\documentclass[a4paper,12pt]{article}
\pdfoutput=1 
\usepackage{graphicx}
\usepackage{bm}
\def\be{\begin{equation}}
\def\ee{\end{equation}}
\def\bea{\begin{eqnarray}}
\def\eea{\end{eqnarray}}
\usepackage{latexsym}
\usepackage{dcolumn}
\usepackage{amsmath}
\usepackage{epsfig,amssymb,euscript,mathrsfs}
\usepackage{array,calc,epsfig}

\usepackage[noadjust]{cite}

\usepackage{chngcntr}

\usepackage[pdfpagelabels]{hyperref}

\usepackage{color}

\renewcommand{\to}{\rightarrow}

\def\be{\begin{equation}}
\def\ee{\end{equation}}
\def\ba{\begin{eqnarray}}
\def\ea{\end{eqnarray}}
\def\nb{\nonumber}

\counterwithin*{equation}{section}

\def\({\left (}
\def\){\right )}
\def\[{\left [}
\def\]{\right ]}

\let\oldlgraf\{ 
\renewcommand{\{}{\left \oldlgraf}

\let\oldrgraf\}
\renewcommand{\}}{\right \oldrgraf}

\usepackage{xfrac} 
\usepackage{float} 
\usepackage{subfig}

\newcommand{\eqnref}[1]{(\ref{#1})}

\newlength\dlf  

\usepackage{booktabs}
\usepackage{comment}
\usepackage{cases} 
\usepackage{empheq}
\usepackage{cancel}

\begin{document}
\baselineskip=15.5pt
\pagestyle{plain}
\setcounter{page}{1}
\newfont{\namefont}{cmr10}
\newfont{\addfont}{cmti7 scaled 1440}
\newfont{\boldmathfont}{cmbx10}
\newfont{\headfontb}{cmbx10 scaled 1728}
\renewcommand{\theequation}{{\rm\thesection.\arabic{equation}}}
\renewcommand{\thefootnote}{\arabic{footnote}}

\vspace{1cm}
\begin{titlepage}
\vskip 2cm
\begin{center}
{\Large{\bf On the Hagedorn Temperature\\ in Holographic Confining Gauge Theories
}}
\end{center}

\vskip 10pt
\begin{center}
Francesco Bigazzi$^{a}$, Tommaso Canneti$^{a,b}$, Aldo L. Cotrone$^{a,b}$
\end{center}
\vskip 10pt
\begin{center}
\vspace{0.2cm}
\textit {$^a$ INFN, Sezione di Firenze; Via G. Sansone 1; I-50019 Sesto Fiorentino (Firenze), Italy.
}\\
\textit{$^b$ Dipartimento di Fisica e Astronomia, Universit\'a di Firenze; Via G. Sansone 1;\\ I-50019 Sesto Fiorentino (Firenze), Italy.
}
\vskip 20pt
{\small{
bigazzi@fi.infn.it, canneti@fi.infn.it, cotrone@fi.infn.it}
}

\end{center}

\vspace{25pt}

\begin{center}
 \textbf{Abstract}
\end{center}

\noindent

The divergence of the string partition function due to the exponential growth of states is a well-understood issue in flat spacetime. It can be interpreted as the appearance of tachyon modes above a certain temperature, known as the Hagedorn temperature $T_H$. In the literature, one can find some intuitions about its generalization to curved spacetimes, where computations are extremely hard and explicit results cannot be provided in general. In this paper, we present a genus-zero estimate of $T_H$, at leading order in $\alpha'$, for string theories on curved backgrounds holographically dual to confining gauge theories. This is a particularly interesting case, since the holographic correspondence equates $T_H$ with the Hagedorn temperature of the dual gauge theories. 
For concreteness we focus on Type IIA string theory on a well known background dual to an $SU(N)$ Yang-Mills theory. The resulting Hagedorn temperature turns out to be proportional to the square root of the Yang-Mills confining string tension. The related coefficient, which at leading order is analytically determined, is the same as the one for Type II theories in flat space. While the calculation is performed in a specific model, the result applies in full generality to confining gauge theories with a top-down holographic dual.

\end{titlepage}

\newpage
\tableofcontents

\section{Introduction}

Theories with a density of states which grows exponentially with the energy are well defined below a certain temperature, called the Hagedorn temperature $T_H$.
The latter is defined as the temperature above which the partition function $Z$ diverges \cite{Hagedorn:1968jf}.
Examples of theories with this behavior include string theories but also ordinary gauge theories, as pure Yang-Mills (YM), where the spectrum of hadrons (glueballs) indeed is believed to grow exponentially.

The computation of the Hagedorn temperature for generic confining theories is not an easy task, even on the lattice, since $T_H$ is larger than the critical temperature for deconfinement (but see e.g.~\cite{Bringoltz:2005xx} for an estimate in pure YM and \cite{Caselle:2015tza} for a discussion).
In this paper we consider confining theories having a dual holographic description with a reliable supergravity regime.
In this case the Hagedorn temperature of the gauge theory is given by the one of the dual string theory on a curved background.

The problem of calculating $T_H$ is well understood for string theories in flat space. The starting point is the asymptotic formula for the number of all partitions of a large-integer $\mathcal{N}$ given by Hardy and Ramanujan \cite{Hardy:1918}. In fact, it can be interpreted as the degeneracy of the $\mathcal{N}$-th level in the spectrum of a one-dimensional bosonic string theory (e.g., see \cite{Zwiebach}). This result has been first generalized in whatever dimension by Huang and Weinberg \cite{Huang:1970iq} and then extended to the superstring case in several works \cite{Sundborg:1984uk,Bowick:1985az,Tye:1985jv,Matsuo:1986es}. The exponential behavior of the resulting density of states is such that $Z$ converges only below a particular limiting temperature, that is $T_H$.

Unfortunately, it is not known how to compute $Z$ for a string theory on a generic curved background. What we can always do is to expand the string action around a classical configuration. The divergence in $Z$ we are looking for seems to originate from temporal winding modes which become tachyonic above $T_H$ \cite{Sathiapalan:1986db,Kogan:1987jd,OBrien:1987kzw}. In this direction, Atick and Witten proposed a genus-zero method for computing $T_H$ in flat space \cite{Atick:1988si}. Their results reproduce the known values of $T_H$ exploiting just the mass-shell condition of the theory, instead of well-known expressions of $Z$ in flat space. 

In this paper, we provide an estimate for the Hagedorn temperature of confining theories with a gravity dual, by applying the Atick-Witten genus-zero method to curved backgrounds, in the Green-Schwarz (GS) formalism.

The estimate of $T_H$ is derived from the semi-classical quantization of the string around the (classical) temporal winding configuration.
The string is placed in the region of the geometry corresponding to the deep infra-red regime of the dual field theory.
The winding mode is supposed to be the lightest state becoming tachyonic as the temperature is increased, providing the value of $T_H$.
The latter is an estimate, rather than the precise value, of the Hagedorn temperature for at least three reasons: because we extrapolate the validity of the semi-classical quantization down to zero mass of the classical configuration; because we cannot completely exclude the (unlikely) possibility that a different configuration becomes tachyonic at a smaller temperature; and because, not being able to fully solve the string theory on curved backgrounds, the result is correct only at leading order in $\alpha'$.  

The mode we consider is the simplest classical configuration (i.e.~it solves the equations of motion and the Virasoro constraints) winding the temporal direction.
It is basically (the Euclidean counter-part of) the configuration providing the field theory string tension $T_s$ in holography or on the lattice.
This elementary observation explains the direct link of $T_H$ to $T_s$.

In this paper we work out the explicit example of the Type IIA string on Witten's background dual to a YM theory coupled to adjoint Kaluza-Klein modes \cite{Witten:1998zw}.
Nevertheless, it will be clear from the computation that the result is completely general in Type II theories dual to confining theories.
The semi-classical quantization of the string provides eight free bosonic modes and their superpartners.\footnote{See \cite{Faraggi:2011ge} for a geometrical approach to this problem.}
We calculate the masses of these modes consistently with the absence of conformal anomaly.
Their values are directly connected to the ones found in the Wilson loop calculation \cite{Bigazzi:2004ze}.

With this spectrum at hand, we derive in the standard way the mass-shell condition, which provides the temperature dependence of the mass of the (quantized) configuration.
The condition of zero mass, giving the Hagedorn temperature, will correspond to vanishing values of the stringy mode masses.
Thus, the calculation reduces to the one in flat space, up to the value of the string tension.

The resulting estimate for the Hagedorn temperature reads 
\begin{equation}\label{result}
\boxed{
T_H = \sqrt{T_s} \sqrt{\frac{1}{4\pi}}
}\,,
\end{equation}
where $T_s$ is the string tension in the dual field theory.
We will also extract a subset of the leading $\alpha'$ corrections to this result. 

For Type II strings in flat space the result is the same as (\ref{result}) with $T_s = 1/2\pi\alpha'$.
In general, for stringy models the expectation is that $T_H/\sqrt{T_s} \sim 1/\sqrt{c_{eff}}$, where $c_{eff}$ is some effective central charge (see e.g.~\cite{Kutasov:1990sv,Sugawara:2012ag}).
Clearly the point is to understand in each case what are the correct values of $T_s$, $c_{eff}$.
Our explicit computation provides evidence to the intuitive expectation that, in Type II theories dual to confining models and at leading order in $\alpha'$, $T_s$ is the dual field theory string tension and $c_{eff}$ is the same as in flat space.

The rest of the paper is organized as follows.
In section \ref{secSetup} we describe the classical configuration winding the temporal direction, which will be the base of our calculation. Section \ref{secQuantization} describes the semi-classical quantization of this configuration and the mass spectra of world-sheet bosonic and fermionic modes. These data are employed to derive the mass-shell condition and, imposing vanishing mass of the quantized mode, the estimate for the Hagedorn temperature in section \ref{secHagedorn}.
We conclude with a few comments in section \ref{secConclusions}.
The appendices contain further technical details of the computations.

{\bf Note added}: after the completion of this work we have realized that, due to subtleties in the light-cone gauge quantization, some statements in this paper are not accurate, although the main results, i.e. the mass-shell condition \eqref{massshellcond} and the leading order expression \eqref{result} for $T_H$, turn out to be correct. We refer to \cite{Bigazzi:2023oqm} for a more rigorous treatment of the problem, in which we explain how to deal with the non-physical (gauge) modes in a proper way and, in particular, we consistently impose the light-cone gauge condition that removes them from the spectrum.

\section{The classical string configuration}
\label{secSetup}

In this section, we aim to describe the string background dual to the so-called \emph{Witten-Yang-Mills (WYM) theory} and the classical configuration that will be quantized in the following. 

WYM is the theory on the world-volume of a stack of $N$ D4-branes wrapped on a circle with anti-periodic boundary conditions for fermions \cite{Witten:1998zw}.
At low energies, it reduces to four-dimensional $SU(N)$ Yang-Mills coupled to massive adjoint Kaluza-Klein (KK) modes.
In the regime where the theory has a reliable supergravity dual, i.e.~the strongly coupled planar (large $N$) limit, the KK modes are at the same mass scale as the glueballs.
The theory displays linear confinement and a mass gap and it is believed to be in the same universality class of pure YM.

The Type IIA supergravity background dual to the WYM theory in the confining phase at finite temperature $T$ is given by
\be \label{WbackOld}
\begin{split}
&ds^2 = \frac{1}{m_0^2} \( \frac{u}{R} \)^{\sfrac{3}{2}} \( \delta_{\mu\nu} dx^\mu dx^\nu + \frac{4}{9} f(u) d\theta^2 \) + \( \frac{R}{u} \)^{\sfrac{3}{2}} \frac{du^2}{f(u)} +R^{\sfrac{3}{2}} u^{\sfrac{1}{2}} d\Omega_4^2 \, , \\
&m_0^2 = \frac{u_0}{R^3} \, , \quad f(u) = 1 - \frac{u_0^3}{u^3} \, , \quad e^{\phi} = g_s \frac{u^{\sfrac{3}{4}}}{R^{\sfrac{3}{4}}} \, , \quad  R = \( \pi N g_s \)^{\sfrac{1}{3}} \alpha'^{\sfrac{1}{2}} \, , \quad F_4 = 3 R^3 \omega_4 \, ,
\end{split}
\ee
where $\mu$, $\nu = 0,1,2,3$ and $\omega_4$ is the volume form of the transverse $S^4$. Notice that, with this notation, the coordinates $x^\mu$ are dimensionless. The above ten-dimensional string frame metric, the dilaton $\phi$ and the constant Ramond-Ramond field strength $F_4$ make up the so-called \emph{Witten background}, taken with Euclidean signature and a time direction which is compactified on a circle of length $1/T$ . Here, $u \in [ u_0, +\infty )$ is the holographic coordinate and $\theta$ is an angular coordinate $\theta \simeq \theta + 2\pi$ which parametrizes a shrinking circle along the $u$-direction. Its radius is asymptotically fixed by the inverse of the glueball and KK mass scale 
\be \label{MKK}
M_{KK} = \frac{3}{2} \, m_0
\ee 
and vanishes at $u=u_0$. 
The value $u=u_0$ corresponds to the position of the tip of the $cigar$ in the $(u, \theta)$-plane. The region $u \sim u_0$ is dual to the IR regime of the WYM theory.

In the following, we will find a classical configuration containing winding modes in the temporal direction and around which we will expand the world-sheet action for type IIA closed strings localized at the tip of the cigar.
The latter requirement comes from the fact that we search for the lighter winding mode which solves the equations of motion.
In order to compute a generalized mass-shell condition, we just need to know the field content of the action and so we have to expand it up to quadratic order in quantum fluctuations. $T_H$ will be deduced fixing to zero the mass of the physical ground state.

The discussion of fermions in a Euclidean background is often problematic. So we will adopt the strategy of Wick-rotating the $x^1$-direction in \eqnref{WbackOld} and work with a Lorentzian background given by
\be \label{Wback}
\begin{split}
&ds^2 = \frac{1}{m_0^2} \( \frac{u}{R} \)^{\sfrac{3}{2}} \( \widetilde\eta_{\mu\nu} dx^\mu dx^\nu + \frac{4}{9} f(u) d\theta^2 \) + \( \frac{R}{u} \)^{\sfrac{3}{2}} \frac{du^2}{f(u)} +R^{\sfrac{3}{2}} u^{\sfrac{1}{2}} d\Omega_4^2 \, , \\
&m_0^2 = \frac{u_0}{R^3} \, , \quad f(u) = 1 - \frac{u_0^3}{u^3} \, , \quad e^{\phi} = g_s \frac{u^{\sfrac{3}{4}}}{R^{\sfrac{3}{4}}} \, , \quad  R = \( \pi N g_s \)^{\sfrac{1}{3}} \alpha'^{\sfrac{1}{2}} \, , \quad F_4 = 3 R^3 \omega_4 \, ,
\end{split}
\ee
where
\be
\widetilde \eta= \text{diag}\{1,-1,1,1\} \, .
\ee
The rotation makes it possible to choose real conjugate momenta. In particular, the one related to $x^1$ will be related to the energy of physical string states living in the nine-dimensional subspace orthogonal to the compact temporal direction.

The spectrum of bosonic excitations does not depend on the Wick rotation, and it can be reliably calculated on the background (\ref{WbackOld}).
Moreover, absence of conformal anomaly links the fermionic mode masses to the bosonic ones.
Thus, we do not expect a direct calculation of the fermionic masses on (\ref{WbackOld}) to give a different result with respect to the one on (\ref{Wback}).
In any case, the final result for the Hagedorn temperature at leading order will not depend on this issue.

In the literature, changing coordinate system as $r^2 \sim u - u_0$ is a very common strategy. Then, one usually introduces locally flat coordinates $y_1$ and $y_2$ such that $r^2=y_1^2+y_2^2$, that is
\begin{subequations}
  \begin{empheq}[left=\empheqlbrace]{align}
    &u = u_0 \, (1+ y_1^2 + y_2^2)  \, , \\
    &\theta =2 \arctan\({\frac{y_1}{\sqrt{y_1^2+y_2^2}+y_2}}\) \, .
  \end{empheq}
\end{subequations}
The background metric \eqnref{Wback} expanded around $y_1=y_2=0$ ($\Rightarrow u=u_0$) reads
\be \label{Wbacktipy}
ds^2 \approx R^{3/2} u_0^{1/2} \{ \[ 1 +\frac{3}{2} \( y_1^2 + y_2^2 \) \] \widetilde\eta_{\mu\nu} dx^\mu dx^\nu + \frac{4}{3} \( dy_1^2 + dy_2^2 
\) + \[ 1 +\frac{1}{2} \( y_1^2 + y_2^2 \) \] d\Omega_4^2 \} \, .
\ee
In this coordinates, it is quite easy to obtain the bosonic mass spectrum (e.g., see \cite{Bigazzi:2004ze}).  Nevertheless, we prefer to work with the coordinates $(u,\theta)$, since, as we will see, this allows us to derive the standard expression for the Hamiltonian of the string in a straightforward way. 

\subsection{The reference classical configuration}
\label{sec:trulyclassconf}

So, let us consider a closed string with Lorentzian flat world-sheet embedded in the Lorentzian background \eqnref{Wback}. Let us fix the world-sheet metric as
\be \label{worldsheetmetric}
d\tilde s^2 = \eta_{\alpha \beta} \, d\sigma^\alpha d\sigma^\beta = -d\tau^2 + d\sigma^2 \, ,
\ee
where $-\infty < \tau < + \infty$ and $0 \leq \sigma < 2\pi$.

In general, the world-sheet bosonic action in its Polyakov form reads 
\be \label{SB}
S_B = -\frac{1}{4 \pi \alpha'} \int d\tau d\sigma \, \eta^{\alpha\beta} \, \partial_{\alpha} \rho^p \, \partial_{\beta} \rho^q \, g_{pq}\(\rho\) \, , \quad \alpha, \beta = \tau, \sigma \, , \quad p, q = 0,...,9 \, ,
\ee
where $g_{pq}$ are the components of the background metric in \eqnref{Wback} and $\rho$ is the set of coordinates 
\be \label{rho}
\rho = \{ x^0, x^1, x^2, x^3, u, \theta, \Omega_4 \} \, . 
\ee

Let us denote the reference classical configuration as $P$. The latter has to be a solution of the equations of motion related to \eqnref{SB} for $\rho=P$, that is
\be \label{classeom}
\eta^{\alpha\beta} \[ \partial_k g_{pq}(P) \, \partial_{\alpha} P^p \, \partial_{\beta} P^q - 2 \partial_\alpha (g_{kq}(P) \, \partial_\beta P^q) \]=0 \, , \quad k = 0, 1, ... , 9 \, .
\ee
Moreover, let us stress that \eqnref{SB} is \emph{classically} equivalent to the Nambu-Goto formulation of the bosonic action if the equations of motion for the world-sheet metric are satisfied. In other words, our classical configuration is subject to the Virasoro constraints
\be \label{Talphabeta0}
T^B_{\alpha\beta} = \partial_\alpha P^p \, \partial_\beta P^q \, g_{pq}(P) - \frac{1}{2} \, \eta_{\alpha\beta} \, \eta^{\rho\sigma} \, \partial_\rho P^p \, \partial_\sigma P^q \, g_{pq}(P) = 0 \, .
\ee
The above conditions turn out to be equivalent to
\be \label{constraints}
A^2 \, h_{\alpha\beta} = \eta_{\alpha\beta} \, ,
\ee
where $A=A(\tau,\sigma)$ is an arbitrary function of the world-sheet coordinates and $h_{\alpha\beta} = \partial_\alpha P^p \, \partial_\beta P^q \, g_{pq}(P)$ is the induced metric on the world-sheet.

Finally, dealing with closed strings at finite temperature, we have to require appropriate periodic boundary conditions in the spatial directions, that is
\be \label{pbc}
P^q(\tau,\sigma+2\pi) = P^q(\tau,\sigma) \, , \quad  q = 1, ... , 9 \, .
\ee
For the temporal direction this will be true up to the length of the circle.

So, let us try to find the minimal classical configuration which contains a temporal winding mode localized at the tip of the cigar, solves the equations of motion \eqnref{classeom}, satisfies the constraints \eqnref{constraints} and respects the periodic boundary conditions \eqnref{pbc}. In the most general case, we have to start from\footnote{Strings orbiting in the internal space correspond to a spectrum of hadronic-like states charged under global flavor symmetries. 
So, we exclude the coordinate of the $S^4$ from the ansatz.}
\be
P = \{ X^0, \, X^i, U, \Theta, 0, 0, 0 , 0 \} \, , \quad i = 1, ..., d \leq 3 \, ,
\ee
for $X^0$, $X^i$, $U$ and $\Theta$ generic functions of $\tau$, $\sigma$. With this notation, the constraints \eqnref{constraints} read
\begin{subequations} \label{Tabeq0}
  \begin{empheq}[left=\empheqlbrace]{align}
    &\label{Tabeq01} \partial_\tau X^\mu \partial_\tau X^\nu \, g_{\mu\nu}(U) + (\partial_\tau U)^2 \, g_{uu}(U) + (\partial_\tau \Theta)^2 \, g_{\theta\theta}(U) = -\frac{1}{A^2} \, ,\\
    &\label{Tabeq02}\partial_\tau X^\mu \, \partial_\sigma X^\nu \, g_{\mu\nu}(U) + \partial_\tau U \, \partial_\sigma U \, g_{uu}(U) + \partial_\tau \Theta \, \partial_\sigma \Theta \, g_{\theta\theta}(U) = 0 \, ,\\
    &\label{Tabeq03} \partial_\sigma X^\mu \partial_\sigma X^\nu \, g_{\mu\nu}(U) + (\partial_\sigma U)^2 \, g_{uu}(U) + (\partial_\sigma \Theta)^2 \, g_{\theta\theta}(U) = \frac{1}{A^2} \, ,
  \end{empheq}
\end{subequations}
and the equations of motion \eqnref{classeom} become
{\footnotesize
{\begin{subequations} \label{classeom2}
  \begin{empheq}[left=\empheqlbrace]{align}
    &\eta^{\alpha\beta} \, \partial_\alpha U \, \partial_\beta X^\mu \, \partial_u g_{00}(U) + g_{00}(U) \, \eta^{\alpha\beta} \, \partial_\alpha \partial_\beta X^\mu = 0 \,  , \quad \mu=0,1,..,d \label{eom1e2} \, ,\\
    &\eta^{\alpha\beta} \, \partial_\alpha U \, \partial_\beta \Theta \, \partial_u g_{\theta\theta}(U) + g_{\theta\theta}(U) \, \eta^{\alpha\beta} \, \partial_\alpha \partial_\beta \Theta = 0 \label{eom3} \, ,\\
    &\eta^{\alpha\beta} \[\partial_\alpha X^\mu \, \partial_\beta X^\nu  \partial_u g_{\mu\nu}(U) + \partial_\alpha U \, \partial_\beta U \partial_u g_{uu}(U) + \partial_\alpha \Theta \, \partial_\beta \Theta \partial_u g_{\theta\theta}(U) - 2 \partial_\alpha( g_{uu}(U) \partial_\beta U) \] = 0 \, . \label{eom4}
  \end{empheq}
\end{subequations}
}}

Let us make some observations. First of all, if we want to include a temporal winding mode in the ansatz, we have to require
\be \label{firstX0}
X^0 = m_0 \frac{\beta}{2\pi} \sigma \, , \quad X^0 \simeq X^0 + m_0\beta \, \quad (\beta\equiv 1/T)\,.
\ee 

Then, if we want to put the string at the tip of the cigar, the first derivatives of $U$ have to vanish in $U=u_0$. Since $g_{uu}(U)$ and its derivatives diverge in $U=u_0$, we have to pay attention in order to avoid indeterminate expressions in \eqnref{Tabeq0} and \eqnref{classeom2}. A reasonable ansatz for $U$ is thus
\be \label{DU}
\partial_\alpha U = \frac{c_\alpha}{(g_{uu}(U))^h} \to 0 \, , \quad U \to u_0 \, , \quad c_\tau, c_\sigma \in \mathbb{R} \, , \quad h \in \mathcal{I} \, ,
\ee
where 
\be
\mathcal{I} = \{ h \in \mathbb{R} \, : h= \frac{1}{2} \lor h \ge 1 \lor \(\frac{1}{2} < h < 1 \land \eta^{\alpha\beta} c_\alpha c_\beta=0 \) \} \, .
\ee
The conditions on $h$ ensure that the constraints \eqnref{Tabeq0} and \eqnref{eom4} are well defined in $U=u_0$, since
\be
\partial_\alpha U \, \partial_\beta U \, \partial_u g_{uu}(U) - 2 \, \partial_\alpha( g_{uu}(U) \partial_\beta U) = (2h-1) \, \partial_\alpha U \, \partial_\beta U \, \partial_u g_{uu}(U)
\ee
and
\be \label{tiplimitDUDUDguu}
\eta^{\alpha\beta}\partial_\alpha U \, \partial_\beta U \partial_u g_{uu}(U) = - \eta^{\alpha\beta} c_\alpha c_\beta \, \frac{3}{2 \, U} \(\frac{U}{R}\)^{3\,h-\frac{3}{2}} \(1+\frac{u_0^3}{U^3}\) \(1-\frac{u_0^3}{U^3}\)^{2(h-1)} \, .
\ee

Furthermore, since $g_{00}(u_0)$, $\partial_u g_{00}(u_0) \ne 0$, \eqref{eom1e2} implies
\be \label{harm}
\eta^{\alpha\beta} \partial_\alpha \partial_\beta X^\mu = 0 \, , \quad \mu=0,1,...,d \,.
\ee
This is a set of necessary conditions if we want to include $U=u_0$ as a part of the ansatz. So, let us choose
\be \label{firstXi}
X^i=\alpha' p^i m_0 \tau \, , \quad p^i \in \mathbb{R} \quad : \quad \widetilde \eta_{ij} \, p^i \, p^j = - M^2 \, , \quad i,j=1, ... , d \, .
\ee
Notice that $M^2$ is the nine-dimensional squared mass of the stringy states under examination. $M^2>0$ for $p^i \in \mathbb{R}$ is possible thanks to the Lorentzian signature of $\widetilde \eta$.  
Moreover, by construction, \eqnref{firstX0} and \eqnref{firstXi} satisfy \eqnref{pbc}.

On the other hand, \eqnref{eom3} does not give any similar information about $\Theta$, since $g_{\theta\theta}(u_0)=0$. Besides, the tip of the cigar corresponds to the origin of the ``polar'' coordinate system and so, in principle, $\Theta$ can take on any possible value. For the sake of simplicity, we assume $\Theta$ to be constant, compatibly with \eqref{pbc}.\footnote{In general, introducing a non-trivial $\Theta=\Theta(\tau,\sigma)$ would make the string modes heavier. So, it is reasonable to focus on constant $\Theta$ from the beginning.} 

Finally, let us note that \eqnref{classeom2} reduces to
\be \label{quantumconstraint}
    (1-2h) \lim_{U\to u_0} \eta^{\alpha\beta}\partial_\alpha U \, \partial_\beta U \, \partial_u g_{uu}(U) = m_0^2 \[\frac{\beta^2}{4\pi^2}+\alpha'^2 M^2\] \partial_u g_{00}(u_0) \, .
\ee
The RHS of the above equation is the sum of non-negative quantities. So, it cannot be satisfied $\forall M^2 >0$ if the LHS is exactly zero. Given \eqnref{tiplimitDUDUDguu}, this excludes all the cases specified in $\mathcal{I}$ except $h=1$.

All in all, let us consider the classical configuration\footnote{Let us stress that $1/g_{uu}(U)$ goes to zero as $U \to u_0$. Therefore, the partial derivatives of $U$ are zero on the classical configuration, as required by the ansatz $U=u_0=constant$.}
\begin{subequations} \label{ansatz}
  \begin{empheq}[left=\empheqlbrace]{align}
    &X^0 = m_0 \frac{\beta}{2\pi} \sigma \, , \quad X^0 \simeq X^0 + m_0\beta \, ,\\
    &\label{Xi}X^i=\alpha' p^i m_0 \tau \, , \quad p^i \in \mathbb{R} \quad : \quad \widetilde \eta_{ij} \, p^i \, p^j = - M^2 \, , \quad i,j=1, ... , d \, ,\\
    &\label{firstU} U=u_0 \quad : \quad  \left . g_{uu}(U) \, \partial_\alpha U \right |_{U=u_0} = c_\alpha \, , \quad c_\tau \, , c_\sigma \in \mathbb{R} \, , \\
    &\label{Theta} \Theta=\text{constant} \, .
  \end{empheq}
\end{subequations}
A similar genus-one version of this configuration has been considered in \cite{PandoZayas:2003jr}.
We can now summarize the classical Virasoro constraint in \eqnref{Tabeq0} and the equations of motion \eqnref{classeom2}, using \eqnref{tiplimitDUDUDguu} and \eqnref{quantumconstraint}, as\footnote{Notice that from \eqnref{Tabeq0} one can read off the coefficients of the induced metric $h_{\alpha\beta}$ on the world-sheet (cfr. with \eqnref{constraints}). In particular, on the classical solution \eqnref{ansatz}, they are constant. This means that the world-sheet has the topology of a cylinder.}
\begin{subequations}
  \begin{empheq}[left=\empheqlbrace]{align}
    &\label{classVirasoro} \alpha'^2 M^2 =  \frac{\beta^2}{4\pi^2} = \frac{1}{m_0^2 \, g_{00}(u_0) A^2} \, ,\\
    &\label{classConstraintOnC}\eta^{\alpha\beta} c_\alpha c_\beta = \frac{1}{2} \[\frac{\beta^2}{4\pi^2}+\alpha'^2 M^2\] \, .
  \end{empheq}
\end{subequations}
En passant, let us stress that \eqnref{classConstraintOnC} selects specific values for the $c_\alpha$-coefficients in \eqnref{firstU} at pure classical level.

Taking the $M^2\to0$ limit of the above equations, we then find
\begin{subequations} \label{classicalconstraint}
  \begin{empheq}[left=\empheqlbrace]{align}
    &\beta_H^{class} = 0 \, ,\\
    &\label{constraint} \left . \eta^{\alpha\beta} c_\alpha c_\beta \right |_{\beta=\beta_H^{class}} = 0 \, ,
  \end{empheq}
\end{subequations}
where $\beta_H^{class}$ is the value of the inverse Hagedorn temperature in the \emph{classical limit} (let us say, superstring theory in curved space neglecting quantum fluctuations), while \eqnref{constraint} represents a constraint which $c_\tau$ and $c_\sigma$ have to satisfy as functions of $\beta$ in this regime.

\section{Semi-classical quantization}
\label{secQuantization}

Having fixed the reference classical configuration, we can return to the action \eqnref{SB} expanding it around \eqnref{ansatz} up to quadratic order in quantum fluctuations. First of all, notice that the classical leading term of such an expansion is expected to be
\be \label{SBcl}
S_B^{cl} = -\frac{1}{4 \pi \alpha'} \int d\tau d\sigma \, \eta^{\alpha\beta} \, \partial_{\alpha} P^p \, \partial_{\beta} P^q \, g_{pq}(P) \, , \quad \alpha, \beta = \tau, \sigma \, , \quad p, q = 0,...,9 \, .
\ee
$P$ still satisfies the equations of motion in \eqnref{classeom2} by construction. On the other hand, the Virasoro constraints have to be written for the whole configuration
\be
\rho = P + \xi \, , 
\ee
including the quantum fluctuations $\xi$. 

At leading order in the low energy, weak coupling limit,
strings appear point-like and all points of the four-sphere are equivalent to each other. Therefore, we can choose to expand the metric around one of these points where the four-sphere looks flat. Introducing proper Cartesian coordinate $\{z^6,z^7,z^8,z^9\}$, we thus have
\be \label{Omegaapprox}
d\Omega_4^2 \approx \delta_{ab} dz^a dz^b \, , \quad a,b = 6,7,8,9 .
\ee

Now, let us focus on each terms of the sum over $p$, $q$ in \eqnref{SB} separately, expanding them around \eqnref{ansatz} up to quadratic order in the quantum fluctuations
\be \label{fluc}
\xi=\{\xi^\mu, \xi^u, \xi^\theta, \xi^a\}\, , \quad \mu=0,1,...,3 \, , \quad a=6,7,8,9 \, .
\ee
Adopting a symbolic but hopefully clear notation, let us start from
\be
\int \partial_\alpha u \, \partial_\beta u \, g_{uu}(u) = \int \partial_\alpha (U+\xi^u) \partial_\beta (U+\xi^u) \[g_{uu}(U) + \partial_u g_{uu}(U) \xi^u + \frac{1}{2} \partial_u^2 g_{uu}(U) (\xi^u)^2\] + o(\xi^2) \, .
\ee
Notice that $\alpha$ and $\beta$ are actually contracted with the inverse world-sheet metric tensor, which is symmetric. So, for the sake of simplicity, let us resume the mixed terms as $\partial_\alpha U \partial_\beta \xi^u + \partial_\alpha \xi^u \partial_\beta U \sim 2 \partial_\alpha U \partial_\beta \xi^u$. In this way, we get
\be \label{DuDuguu}
\begin{split}
\int \partial_\alpha u \, \partial_\beta u \, g_{uu}(u) =
&\int \partial_\alpha U \partial_\beta U  g_{uu}(U) + \partial_\alpha \xi^u \partial_\beta \xi^u  g_{uu}(U) + \frac{1}{2} \partial_\alpha U \partial_\beta U \partial_u^2 g_{uu}(U) (\xi^u)^2 + \\
& \hspace{-1cm} +\[\partial_\alpha U \partial_\beta U \partial_u g_{uu}(U) + 2 \partial_\alpha U \partial_\beta \xi^u \partial_u g_{uu}(U)\] \xi^u + 2 \partial_\alpha U \partial_\beta \xi^u  g_{uu}(U) + o(\xi^2) \, .
\end{split}
\ee
Integrating by parts the last two terms (setting to zero the boundary terms), we obtain
\be \label{DuDuguudef}
\begin{split}
\int \partial_\alpha u \, \partial_\beta u \, g_{uu}(u) =
&\int \partial_\alpha U \partial_\beta U  g_{uu}(U) + \partial_\alpha \xi^u \partial_\beta \xi^u  g_{uu}(U) + \\
&-\[\partial_\alpha U \partial_\beta U \partial_u g_{uu}(U) + 2 \, \partial_\alpha \partial_\beta U  g_{uu}(U)\] \xi^u + \\
&- \frac{1}{2} \[ \partial_\alpha U \partial_\beta U \, \partial_u^2 g_{uu}(U) + 2 \, \partial_\alpha \partial_\beta U \partial_u g_{uu}(U) \]  (\xi^u)^2 + o(\xi^2) \, .
\end{split}
\ee
Then, let us evaluate
\be
\int \partial_\alpha \theta \, \partial_\beta \theta \, g_{\theta\theta}(u) 
= \int \partial_\alpha (\Theta+\xi^\theta) \partial_\beta (\Theta+\xi^\theta) \[g_{\theta\theta}(U) + \partial_u g_{\theta\theta}(U) \xi^u + \frac{1}{2} \partial_u^2 g_{\theta\theta}(U) (\xi^u)^2\] + o(\xi^2) \, .
\ee
that is
\be \label{DthetaDthetagthetatheta}
\int \partial_\alpha \theta \, \partial_\beta \theta \, g_{\theta\theta}(u)= \int \partial_\alpha \xi^\theta \partial_\beta \xi^\theta  g_{\theta\theta}(U) + o(\xi^2) \, .
\ee
Finally, let us consider
{\small{
\be
\begin{split}
 \hspace{-12pt} \int \hspace{-3pt} \partial_\alpha x^\kappa \, \partial_\beta x^\gamma \hspace{-2pt} g_{\kappa\gamma}(u) &  =
\hspace{-5pt} \int \hspace{-5pt} \partial_\alpha (X^\kappa+\xi^\kappa) \partial_\beta (X^\gamma+\xi^\gamma) \[g_{\kappa\gamma}(U) + \partial_u g_{\kappa\gamma}(U) \xi^u + \frac{1}{2} \partial_u^2 g_{\kappa\gamma}(U) (\xi^u)^2\] + o(\xi^2) \\
&  = \hspace{-5pt} \int \hspace{-5pt} \partial_\alpha X^\kappa \partial_\beta X^\gamma  g_{\kappa\gamma}(U) + \partial_\alpha \xi^\kappa \partial_\beta \xi^\gamma  g_{\kappa\gamma}(U) + \frac{1}{2} \partial_\alpha X^\kappa \partial_\beta X^\gamma \partial_u^2 g_{\kappa\gamma}(U) (\xi^u)^2 + \\
&  +  \[\partial_\alpha X^\kappa \partial_\beta X^\gamma \partial_u g_{\kappa\gamma}(U) + 2 \partial_\alpha X^\kappa \partial_\beta \xi^\gamma \partial_u g_{\kappa\gamma}(U)\] \xi^u + 2 \partial_\alpha X^\kappa \partial_\beta \xi^\gamma  g_{\kappa\gamma}(U) + o(\xi^2)\, .
\end{split}
\ee
}}
Integrating by parts the last term, we get
\be \label{dxdxg00}
\begin{split}
\int \partial_\alpha x^\mu \, \partial_\beta x^\nu \, g_{\mu\nu}(u) =
&\int \partial_\alpha X^\kappa \partial_\beta X^\gamma  g_{\kappa\gamma}(U) + \partial_\alpha \xi^\mu \partial_\beta \xi^\nu  g_{\mu\nu}(U) + \\
&-2\[\partial_\alpha X^\kappa \partial_\beta U \partial_u g_{\kappa\gamma}(U) + \partial_\alpha \partial_\beta X^\kappa  g_{\kappa\gamma}(U)\] \xi^\gamma + \partial_\alpha X^\kappa \partial_\beta X^\gamma \partial_u g_{\kappa\gamma}(U) \xi^u \\
&+2 \partial_\alpha X^\kappa \partial_u g_{\kappa\gamma}(U) J_\beta^{u\gamma} + \frac{1}{2} \partial_\alpha X^\kappa \partial_\beta X^\gamma \partial_u^2 g_{\kappa\gamma}(U)  (\xi^u)^2 + o(\xi^2) \, ,
\end{split}
\ee 
where $\mu$, $\nu=0,1,...,3$, $\kappa$, $\gamma=0,1,...,d$, and
\be
J_\beta^{u\gamma} = \xi^u \partial_\beta \xi^\gamma \, .
\ee

Collecting each contribution in \eqnref{DuDuguudef}, \eqnref{DthetaDthetagthetatheta} and \eqnref{dxdxg00}, the coefficients of the $o(\xi)$-terms reproduce the left hand sides of the equations of motion \eqnref{classeom2} and then they can be set to zero. Therefore, the bosonic quadratic action reads
\be \label{SB2}
S_B^{(2)} \hspace{-4pt}= S_B^{cl} - \frac{1}{4 \pi \alpha'} \hspace{-3pt} \int \hspace{-4pt} d\tau d\sigma \, \eta^{\alpha\beta} \[ \partial_\alpha \xi^p \partial_\beta \xi^q g_{pq}(U) + 2 \, \partial_\alpha X^\kappa J_\beta^{u\gamma} \partial_u g_{\kappa\gamma}(U) +  m_{\alpha\beta}^2 (\xi^u)^2  \] ,
\ee
where $S_B^{cl}$ is defined in \eqnref{SBcl}, $p$, $q=0,1,...,9$, $\kappa$, $\gamma=0,1,...,d$ and
\be
m_{\alpha\beta}^2 = \frac{1}{2} \[ \partial_\alpha X^\kappa \partial_\beta X^\gamma \partial_u^2 g_{\kappa\gamma}(U) - \partial_\alpha U \partial_\beta U \partial_u^2 g_{uu}(U) - 2 \partial_\alpha \partial_\beta U \partial_u g_{uu}(U) \] \, .
\ee

In order to interpret the quantum fluctuations as physical matter fields and read off the related mass terms, we have to rescale them in such a way that the kinetic terms are canonically normalized. So, let us introduce the \emph{physical} fluctuations
\be \label{rescaling}
\widetilde \xi^p = \sqrt{g_{pp}(U)} \,  \xi^p \, , \quad p=0,1,...,9 \, .
\ee
In this way,
\be
\partial_\alpha \xi^p = \frac{1}{\sqrt{g_{pp}(U)}} \[ - \frac{1}{2} \, \partial_\alpha U \, \partial_u \ln g_{pp}(U) \, \widetilde \xi^p + \partial_\alpha \widetilde \xi^p \] \, ,
\ee
from which\footnote{As usual, the contraction with $\eta^{\alpha\beta}$ is understood and so we can sum the mixed terms of the product together.}
\be
\int \partial_\alpha \xi^p \partial_\beta \xi^p g_{pp}(U) = \int \partial_\alpha \widetilde \xi^p \partial_\beta \widetilde \xi^p + \frac{1}{4} \partial_\alpha U \partial_\beta U (\partial_u \ln g_{pp}(U))^2 (\widetilde \xi^p)^2- \partial_\alpha U \partial_u \ln g_{pp}(U) \widetilde \xi^p \partial_\beta \widetilde \xi^p \, .
\ee
Integrating by parts the last term, we get
\be
\int d\tau d\sigma \, \eta^{\alpha\beta} \, \partial_\alpha \xi^p \, \partial_\beta \xi^p \, g_{pp}(U) = \int d\tau d\sigma \{ \eta^{\alpha\beta} \, \partial_\alpha \widetilde \xi^p \, \partial_\beta \widetilde \xi^p + M_p^2 \, (\widetilde \xi^p)^2 \} \, , \quad \forall p=0,1,...,9 \, ,
\ee
where
\be
M_p^2 = \frac{1}{2} \eta^{\alpha\beta} \[\partial_\alpha \partial_\beta U \partial_u \ln g_{pp}(U) + \frac{1}{2} \partial_\alpha U \partial_\beta U (\partial_u \ln g_{pp}(U))^2 + \partial_\alpha U \partial_\beta U \partial_u^2 \ln g_{pp}(U)\] \, .
\ee
All in all, the quadratic bosonic action written in terms of physical fluctuations reads
\be \label{SB3}
\begin{split}
S_B^{(2)} = S_B^{cl} - \frac{1}{4 \pi \alpha'} \int  d\tau d\sigma & \{ \eta^{\alpha\beta} (\partial_\alpha \widetilde \xi^p \partial_\beta \widetilde \xi^q \delta_{pq} + 2 \partial_\alpha X^\kappa \partial_u g_{\kappa\gamma}(U) J_\beta^{u\gamma} \right . \\
& \quad \, \left . + \mathcal{M}_\mu^2 \, \widetilde\xi^\mu \, \widetilde\xi^\nu \delta_{\mu\nu} + \mathcal{M}_u^2 \, (\widetilde \xi^u)^2 + \mathcal{M}_\theta^2 \, (\widetilde \xi^\theta)^2 + \mathcal{M}_z^2 \, \widetilde\xi^{a} \, \widetilde\xi^{b} \, \delta_{ab}  \} \, ,
\end{split}
\ee
where $p$, $q=0,1,...,9$, $\kappa$, $\gamma=0,1,...,d$, $\mu$, $\nu=0,1,...,3$, $a$, $b=6,...,9$ and
\begin{subequations}
\begin{align}
&J_\beta^{u\gamma} = \frac{1}{\sqrt{g_{uu}(U)g_{00}(U)}} \[ - \frac{1}{2} \partial_\beta U \partial_u \ln g_{00}(U) \, \widetilde\xi^u \, \widetilde \xi^\gamma + \widetilde\xi^u \, \partial_\beta \widetilde \xi^\gamma \] \, ,\\
&\mathcal{M}_\mu^2 = M_\mu^2 \, , \quad \mathcal{M}_\theta^2 = M_\theta^2  \, ,\quad \mathcal{M}_z^2 = M_z^2 \, , \\
&\mathcal{M}_u^2 = M_u^2 + \frac{\eta^{\alpha\beta}m_{\alpha\beta}^2}{g_{uu}(U)} \, .
\end{align}
\end{subequations}
Plugging the ansatz \eqnref{ansatz} for generic $U$ in the above masses, we would get
\begin{subequations} \label{masses}
\begin{align}
&&\hspace{-24pt}\mathcal{M}_\mu^2 = & \frac{3}{16 U^2} \, \eta^{\alpha\beta} c_\alpha c_\beta \( \frac{U}{R} \)^{3} \(1-\frac{u_0^3}{U^3}\) \( 5 + 7 \frac{u_0^3}{U^3} \) \, , \\
&&\hspace{-24pt}\mathcal{M}_u^2 = &\frac{9}{16 U^2} \, \eta^{\alpha\beta} c_\alpha c_\beta \( \frac{U}{R} \)^{3} \(1+\frac{u_0^3}{U^3}\)^{2} + \frac{3}{8} m_0^2 \[\frac{\beta^2}{4\pi^2}+\alpha'^2 M^2\] \frac{U}{u_0} \(1-\frac{u_0^3}{U^3}\)\,,\\
&&\hspace{-24pt}\mathcal{M}_\theta^2 = & \frac{3}{16 U^2} \, \eta^{\alpha\beta} c_\alpha c_\beta \( \frac{U}{R} \)^{3} \( 5 - 6 \frac{u_0^3}{U^3} + 13 \frac{u_0^6}{U^6} \) , \\
&&\hspace{-24pt}\mathcal{M}_z^2  = & \frac{3}{16 U^2} \, \eta^{\alpha\beta} c_\alpha c_\beta \( \frac{U}{R} \)^{3} \(1-\frac{u_0^3}{U^3}\) \( 1+ 3 \frac{u_0^3}{U^3} \) \, .
\end{align}
\end{subequations}
So, the quadratic bosonic action would describe the dynamics of interacting massive scalar fields. Nevertheless, we are interested in closed strings located at the tip of the cigar, such that
\be
J_\beta^{u\gamma} \to 0 \, , \quad U \to u_0 \, , \quad \forall \gamma = 0,1,...,d \, .
\ee
Therefore $\widetilde\xi^u$ and $\widetilde\xi^\gamma$ decouple in a straightforward way and we get a theory of free scalar fields on a two-dimensional flat world-sheet. 

In the $U \to u_0$ limit, the masses in \eqnref{masses} become
\be \label{MuMtheta}
\mathcal{M}_\mu^2 , \mathcal{M}_z^2 \to 0 \, , \qquad \mathcal{M}_u^2 , \mathcal{M}_\theta^2 \to \frac{9}{4} m_0^2 \, \eta^{\alpha\beta} c_\alpha c_\beta \equiv M_B^2 \,.
\ee
As mentioned above, we can use \eqnref{classConstraintOnC} to express the $c_\alpha$-coefficients in terms of $\beta$. In particular, it follows that
\be
M_B^2 = \frac{9}{8} m_0^2 \[\frac{\beta^2}{4\pi^2}+\alpha'^2 M^2\] \, .
\ee
Let us now remember that we have read off $M_B^2$ as the tip limit of the coefficient of the fluctuations $\left(\widetilde \xi^u\right)^2$ or $\left(\widetilde \xi^\theta\right)^2$ in the bosonic quadratic action \eqnref{SB3}. Thus, by construction, $M_B$ has to be a zero-order coefficient in the expansion. This implies that we have to evaluate it using the classical Virasoro constraint \eqnref{classVirasoro}. Hence
\be \label{MB}
M_B^2 = \frac{9}{4} m_0^2 \frac{\beta^2}{4\pi^2} = \frac{M_{KK}^2}{4\pi^2} \beta^2 \, .
\ee

Light-Cone Gauge quantization teaches us that the physical string degrees of freedom correspond to the transverse oscillators w.r.t.~the two spacetime light-cone coordinates. Therefore, the field content of \eqnref{SB3} in the tip limit is reduced to six massless propagating modes plus two massive propagating modes of mass $M_B$. Moreover, a consistent theory without conformal anomaly has to satisfy the mass matching condition\footnote{In general, this condition may not ensure the absence of conformal anomaly. If the world-sheet has a non-trivial topology, then other curvature-dependent terms arise (e.g., see \cite{Gautason:2021vfc}).}
\be \label{massmatchingcondition}
\sum_{\substack{B \\bosons}} M_B^2 = \sum_{\substack{F \\fermions}} M_F^2 \, ,
\ee
that is the sum of the squared masses of the bosonic propagating modes has to be equal to the sum of the squared masses of the fermionic propagating modes (see \cite{Bigazzi:2004ze} and references therein). 
Let us now turn our attention to the fermionic sector. 

\subsection{Fermionic modes}

To fix the notations, we use greek indices as two-dimensional world-sheet indices, underlined latin indices to denote ten-dimensional flat indices and latin indices for the generic curved ones. The last two are related by a proper set of vielbein $e^{\underline a}$ such that
\be
e^{\underline a}_p \, e^{\underline b}_q \, \widetilde \eta_{\underline a \underline b} = g_{pq}(u) \, , \quad \widetilde \eta=\text{diag}\{1,-1,1,1,1,1,1,1,1,1\} \, .
\ee
The Type IIA GS action on a Lorentzian background expanded up to the second order in the ten dimensional Majorana spinor $\psi=\psi(\tau,\sigma)$ reads \cite{Martucci:2003gc}
\be \label{SF}
S_{F}^{\(2\)} = \frac{i}{4 \pi \alpha'} \int d\tau \, d\sigma \, \bar \psi \[ \( \eta^{\alpha\beta} + \varepsilon^{\alpha\beta} \Gamma^{\underline{11}} \) \Gamma_\alpha D_\beta \] \psi \, ,
\ee
where $\Gamma_\alpha$ is the pull-back on the world-sheet of the Minkowskian gamma matrix $\Gamma_p = e^{\underline a}_{\, . \, p} \, \Gamma_{\underline a}$, that is 
\be \label{pullbackGammagen}
\Gamma_\alpha = \partial_{\alpha} \rho^p \, \Gamma_p
= \partial_{\alpha} \rho^p \, e^{\underline a}_{\, . \, p} \, \Gamma_{\underline a} \, ;
\ee
$\Gamma^{\underline{11}}=\Gamma^{\underline{0}} \cdot ... \cdot \Gamma^{\underline{9}}$ is the ten-dimensional chirality operator; $\varepsilon^{\alpha\beta}$ is the two-dimensional Levi-Civita symbol such that $\varepsilon^{\tau\sigma}=+1$; $\bar \psi = \psi^\dagger \, \Gamma^{\underline 1}$ as usual;\footnote{Notice that, in our conventions, $\Gamma^{\underline 1}$ is the only time-like gamma matrix.} $D_\beta$ is the pull-back on the world-sheet of the generalized covariant derivative
\be \label{D}
D_p = \partial_p + \frac{1}{4} \omega_{\underline a \underline b, \, p} \Gamma^{\underline a \underline b} - \frac{1}{8 \cdot 4! \, g_s} e^{\phi} F^{(4)}_{\underline a \underline b \underline c \underline d} \Gamma^{\underline a \underline b \underline c \underline d} \Gamma_p \, ,
\ee
where the spin-connection $\omega$ is defined as\footnote{As a remark, let us stress that here $\Gamma^{v}_{pq}$, $v$, $p$, $q = 0, ..., 9$, denote the Christoffel symbols related to the metric \eqnref{Wback}.
}
\be \label{spinconn}
\omega^{\underline c}_{\, . \, \underline b, \,p} = e^{\underline c}_{. \, q} \, \partial_p e_{\underline b}^{\, . \, q} + e_{\underline b}^{\, . \, q} \, e^{\underline c}_{. \, r} \, \Gamma^{r}_{pq} \, ,
\ee
and we have adopted the notation
\be
\Gamma_{s_1, ... , s_n} =  \frac{1}{n!} \sum_{\pi} \text{sgn}\(\pi\) \Gamma_{s_{\pi\(1\)}} \cdot ... \cdot \Gamma_{s_{\pi\(n\)}} \, ,
\ee
where the sum runs over all the possible permutations $\pi$ of the set $\{1, ... , n\}$.

Now we have to compute \eqnref{pullbackGammagen} and the pull-back on the world-sheet of \eqnref{D}. Note that \eqnref{SF} is already quadratic in the fermionic fields. So, we have to make explicit these quantities without introducing further field dependence. Basically, every function of the coordinates has to be evaluated \emph{exactly} at classical level, that is at leading order in the expansion around \eqnref{ansatz}. 

We collect some technical details of the computation in appendix \ref{appA}.
In particular, given the set of vielbein reported in \eqnref{vielbein}, \eqnref{pullbackGammagen} reduces to\footnote{We adopt a synthetic notation denoting $\partial_{\alpha} X^\mu \delta^{\underline \mu}_{. \, \mu}  \Gamma_{\underline \mu}$ as $ \partial_{\alpha} X^\mu \,  \Gamma_{\underline \mu}$.}
\be \label{pullbackGamma}
\Gamma_\alpha^{\text{tip}} = \sqrt{|g_{00}(u_0)|} \, \partial_{\alpha} X^\mu \,  \Gamma_{\underline \mu} = \sqrt{m_0 R^3} \, \partial_\alpha X^\mu \,  \Gamma_{\underline \mu} \, , 
\ee
while in appendix \ref{appD} we derive that
\be
D_\alpha^{\text{tip}} = \partial_\alpha - \frac{3}{8} \, \partial_{\alpha} X^\mu \, \widetilde\Gamma \, \Gamma_{\underline \mu} \, , \quad \widetilde\Gamma = {\Gamma}^{\underline 6 \underline 7 \underline 8 \underline 9} \, ,
\ee
where $ X^\mu$ are the classical $\mu$-components of $\rho$, for $\mu=0,1,...,d$.

All in all, the quadratic fermionic action reads\footnote{\label{rescale1}We have also rescaled the Majorana spinor as $\psi \mapsto \( m_0 R^3 \)^{-\sfrac{1}{4}} \psi$.}
\be \label{ESF2}
S_{F}^{\(2\)} = \frac{i}{4 \pi \alpha'} \int d\tau \, d\sigma \, \bar \psi \[ \( \eta^{\alpha\beta} + \varepsilon^{\alpha\beta} \Gamma^{\underline{11}} \) \partial_\alpha X^\mu \,  \Gamma_{\underline \mu} \(\partial_\beta - \frac{3}{8} \partial_{\beta} X^\nu \, \widetilde\Gamma \, \Gamma_{\underline \nu}\) \] \psi \, .
\ee
Notice that $\Gamma_{\underline a}$, $a=0,...,9$, form a set of Lorentzian gamma matrices such that \cite{VanProeyen:1999ni}
\be \label{Clifford}
\begin{split}
& \hspace{-4pt} \{ \Gamma_{\underline a} \, , \Gamma_{\underline b} \} = 2 \, \widetilde\eta_{\underline a \underline b} \, \mathbb{I}_{32} \, , \quad \{ \Gamma_{\underline a} \, , \Gamma^{\underline{11}} \} = 0 \, , \quad \underline a \, , \underline b = \underline 0, ..., \underline 9 \,,\\ 
&\Gamma_{\underline 0}^{\dagger} = \Gamma_{\underline 0} \, , \quad \Gamma_{\underline 1}^{\dagger} = - \Gamma_{\underline 1}\, , \quad \Gamma_{\underline c}^{\dagger} = \Gamma_{\underline c} \, , \quad c=2,...,9 \, , \quad \Gamma^{\underline{11} \, \dagger} = \Gamma^{\underline{11}} \, .
\end{split}
\ee
Now, we can decompose $\psi$ as
\be
\psi = \psi_1 + \psi_2 \, , \quad \psi_I = \frac{1}{2} \( \mathbb{I}_{32} + \(-1 \)^I \, \Gamma^{\underline{11}} \) \psi \, .
\ee
Exploiting only the defining properties in \eqnref{Clifford}, $\psi_1$ and $\psi_2$ turn out to be orthogonal to each other and to be eigenvector of $\Gamma^{\underline{11}}$, that is
\be \label{chiral}
\psi^\dagger_1 \psi_2 = \psi^\dagger_2 \psi_1 = 0 \, , \quad \Gamma^{\underline{11}} \, \psi_I = (-1)^I \, \psi_I \, .
\ee

Then, the GS action has an additional local fermionic symmetry, the so-called \emph{kappa-symmetry}. As a consequence, half of the components of $\psi$ are actually decoupled from the theory \cite{Green:2012oqa}. Let us fix such a symmetry as
\be
\begin{cases}
\Gamma^- \psi_1 = 0\\
\Gamma^+ \psi_2 =0
\end{cases} ,
\ee
where $\Gamma^\pm$ are two different linear combinations of $\Gamma_{\underline 0}$ and $\Gamma_{\underline 1}$, for instance $\Gamma^\pm = \Gamma^{\underline 0} \pm \Gamma^{\underline 1}$. The above conditions can be multiplied by $\Gamma_{\underline 0}$ and rewritten as
\be \label{gaugefixing}
\Gamma_{\underline 0} \Gamma_{\underline 1} \, \psi_I = (-1)^I \, \psi_I \, .
\ee

To fix the ideas, a particular representation of the Lorentzian gamma matrices which satisfies both \eqnref{Clifford} and the gauge fixing \eqnref{gaugefixing} is
\be \label{rep}
\Gamma_{\underline 0} = \sigma_1 \otimes \mathbb{I}_{16} \, , \quad \Gamma_{\underline 1} = i \, \sigma_2 \otimes \mathbb{I}_{16} \, , \quad \Gamma_{\underline A} = \sigma_3 \otimes \gamma_{\underline A} \,, \quad {\underline A} = 2, ... , 9 \, ,
\ee
where $\sigma_1 = \begin{pmatrix} 0 & 1 \\ 1 & 0 \end{pmatrix}$, $\sigma_2 = \begin{pmatrix} 0 & -i \\ i & 0 \end{pmatrix}$ and $\sigma_3 = \begin{pmatrix} 1 & 0 \\ 0 & -1 \end{pmatrix}$ are the Pauli matrices and $\gamma_{\underline A}$ are Euclidean Dirac matrices in eight dimensions, such that $\{\gamma_{\underline A}, \gamma_{\underline B}\}=2 \, \mathbb{I}_{16} \, \delta_{\underline A \underline B}$, $\underline A$, $\underline B = 2, ..., 9$. They can be expressed as
\be \label{gammaA}
\gamma_{\underline A} = \begin{pmatrix} 0 & \Lambda_{\underline A} \\ \Lambda_{\underline A}^T & 0 \end{pmatrix} \, , \quad \Lambda_{\underline A} \Lambda_{\underline B}^T + \Lambda_{\underline B} \Lambda_{\underline A}^T = 2 \, \mathbb{I}_{8} \, \delta_{\underline A \underline B} \, , \quad \underline A, \underline B = 2, ..., 9 \, ,
\ee
where \cite{Green:2012oqa}
\begin{align} \label{Lambda}
&\Lambda_2 = \sigma_1 \otimes i \sigma_2 \otimes \mathbb{I}_2\,, & &\Lambda_6 = i \sigma_2 \otimes i \sigma_2 \otimes i \sigma_2\,, \nb \\
&\Lambda_3 = \sigma_3 \otimes i \sigma_2 \otimes \mathbb{I}_2\,, & &\Lambda_7 = \mathbb{I}_2 \otimes \sigma_1 \otimes i \sigma_2\,, \nb \\
&\Lambda_4 = i \sigma_2 \otimes \mathbb{I}_2 \otimes \sigma_3\,, & &\Lambda_8 = \mathbb{I}_2 \otimes \sigma_3 \otimes i \sigma_2\,, \nb \\
&\Lambda_5 = \mathbb{I}_2 \otimes \mathbb{I}_2 \otimes \mathbb{I}_2\,, & &\Lambda_8 = i \sigma_2 \otimes \mathbb{I}_2 \otimes \sigma_1 \, .
\end{align}
With this choice we have
\be \label{gamma0gamma1}
\Gamma_{\underline 0} \Gamma_{\underline 1} = \begin{pmatrix} -\mathbb{I}_{16} & 0 \\ 0 & +\mathbb{I}_{16} \end{pmatrix} \, , \quad \Gamma^{\underline{11}} = - \begin{pmatrix} -\mathbb{I}_{16} & 0 \\ 0 & +\mathbb{I}_{16} \end{pmatrix} \begin{pmatrix} \gamma^{\underline{11}} & 0 \\ 0 & \gamma^{\underline{11}} \end{pmatrix} \, ,
\ee
where
\be \label{8dgamma11}
\gamma^{\underline{11}} = \gamma_{\underline 2} \cdot ... \cdot \gamma_{\underline 9} = \begin{pmatrix} + \, \mathbb{I}_{8} & 0 \\ 0 & - \, \mathbb{I}_{8} \end{pmatrix}
\ee
is the eight-dimensional chirality operator. Therefore, we have to express $\psi_1$ and $\psi_2$ as
\be
\psi_1 = \begin{pmatrix} \Psi_1 \\ 0 \end{pmatrix} \, , \quad \psi_2 =  \begin{pmatrix} 0 \\ \Psi_2 \end{pmatrix} \quad s. t. \quad \gamma^{\underline{11}} \Psi_I = - \, \Psi_I \, , \, I=1,2 \,,
\ee
in order to simultaneously satisfy the defining property \eqnref{chiral} and the gauge-fixing condition \eqnref{gaugefixing}. In other words, $\Psi_1$ and $\Psi_2$ have to be two eight-dimensional Majorana-Weyl spinors of the same (negative) chirality w.r.t. $\gamma^{\underline{11}}$. Looking at \eqnref{8dgamma11}, it follows that
\be
\Psi_I = \begin{pmatrix} 0 \\ \zeta_I \end{pmatrix} \, , \quad I=1,2 ,
\ee
being $\zeta_1$ and $\zeta_2$ two six-dimensional Majorana spinors. Therefore, the non-vanishing components of $\psi$ correspond to the 8+8=16 real components of $\zeta_1$ and $\zeta_2$. This shows how the gauge-fixing of the kappa-symmetry halves the total fermionic degrees of freedom.

Let us stress that we have used $\widetilde \Gamma$ to denote the antisymmetrization of the product $\Gamma_{\underline 6} \, \Gamma_{\underline 7} \, \Gamma_{\underline 8} \, \Gamma_{\underline 9}$. Nevertheless, the anticommutation relations in \eqnref{Clifford} tell us that such a product is already antisymmetric in ${\underline 6} \leftrightarrow {\underline 7} \leftrightarrow {\underline 8} \leftrightarrow {\underline 9}$. So, it follows that $\widetilde \Gamma = \Gamma_{\underline 6} \, \Gamma_{\underline 7} \, \Gamma_{\underline 8} \, \Gamma_{\underline 9}$. Moreover, given the representation \eqnref{rep}, \eqnref{gammaA}, \eqnref{Lambda}, one can check that
\begin{subequations} \label{gammaproducts}
\begin{align}
& \psi^\dagger \, \Gamma_{\underline 1} \, \Gamma_{\underline i} \, \Gamma^{\underline{11}} \, \overline \Gamma \, \psi = 0 \, , \quad i=1,...,d \, , \quad \forall \psi^T = \begin{pmatrix} 0 & \zeta_1 & 0 & \zeta_2 \end{pmatrix} \, , \\
& \psi^\dagger \, \Gamma_{\underline 1} \, \Gamma_{\underline i} \, \partial_\alpha \psi = 0 \, , \quad i=2,...,d \, , \quad \alpha=\tau, \,\sigma \, , \quad \forall \psi^T = \begin{pmatrix} 0 & \zeta_1 & 0 & \zeta_2 \end{pmatrix} \, , \\
& \psi^\dagger \, \Gamma_{\underline 1} \, \Gamma_{\underline i} \, \Gamma^{\underline{11}} \, \partial_\alpha \psi = 0 \, , \quad i=2,...,d \, , \quad \alpha=\tau, \,\sigma \, , \quad \forall \psi^T = \begin{pmatrix} 0 & \zeta_1 & 0 & \zeta_2 \end{pmatrix} \, ,
\end{align}
\end{subequations}
where
\be \label{specificGamma}
\overline \Gamma \equiv \Gamma_{\underline 1} \, \widetilde \Gamma = i \, \sigma_2 \otimes \widetilde \gamma \, , \quad \( \widetilde \gamma \)_{ij} = (-1)^{i+1} \delta_{ij} \, , \, i,j=1,...,16 \, .
\ee
So, \eqnref{ESF2} reduces to
\be \label{SF2current}
S_F^{(2)}=-\frac{i}{4\pi\alpha'} \int d\tau \, d\sigma \, \psi^\dagger \{ - m_0 \,\frac{\beta}{2\pi} \, \Gamma_{\underline 0} \Gamma_{\underline 1} \, G(p^1) \, (\Gamma^{\underline{11}} \, \partial_\tau + \partial_\sigma) \, \psi - \frac{1}{3} M_B^2 \, H(p^1) \, \overline \Gamma \, \psi \} \, ,
\ee
where
\be
G(p^1) = \mathbb{I}_{32} - \, \frac{2\pi}{\beta}\, \alpha' \, p^1 \, \Gamma_{\underline 0} \, \Gamma_{\underline 1} \, \Gamma^{\underline{11}} \, , \quad H(p^1) = \mathbb{I}_{32} - \frac{\frac{4\pi}{\beta}\alpha' p^1}{1+\frac{4\pi^2}{\beta^2}\alpha'^2 M^2} \Gamma_{\underline 0} \, \Gamma_{\underline 1} \, \Gamma^{\underline{11}} \, .
\ee

Now, exploiting \eqnref{chiral}, \eqnref{gaugefixing} and \eqnref{specificGamma} we can write the fermionic action in terms of the (Grassmann) real components $\zeta_I^\kappa$, $I=1$, $2$, $\kappa=1,...,8$, as
{\small{
\be \label{SF2components}
S_F^{(2)} \hspace{-2pt}=\hspace{-2pt}\frac{i}{4\pi\alpha'} \hspace{-4pt} \int \hspace{-4pt} d\tau \, d\sigma \hspace{-2pt} \sum_{\kappa=1}^8 \hspace{-2pt} \{ m_0 \frac{\beta}{2\pi} \, \widetilde G(p^1) [-\zeta_1^\kappa ( - \partial_\tau + \partial_\sigma) \, \zeta_1^\kappa+\zeta_2^\kappa (\partial_\tau + \partial_\sigma) \, \zeta_2^\kappa] + \frac{2}{3} M_B^2 \, \widetilde H(p^1) \, (-1)^{\kappa} \zeta_{2}^k \zeta_{1}^k \} ,
\ee
}}where
\be \label{GHtilde}
\widetilde G(p^1) = 1 - \, \frac{2\pi}{\beta}\, \alpha' \, p^1 \, , \quad \widetilde H(p^1) = 1 - \frac{\frac{4\pi}{\beta}\alpha' p^1}{1+\frac{4\pi^2}{\beta^2}\alpha'^2 M^2}  \, .
\ee
The equations of motion related to the above action are
\be \label{eom}
\begin{cases}
\( - \partial_\tau + \, \partial_\sigma \) \zeta_1^\kappa  = (-1)^{\kappa+1} \, M_F \, \zeta_2^\kappa\\
\( \, \partial_\tau + \, \partial_\sigma \) \zeta_2^\kappa  = (-1)^{\kappa+1} \, M_F \, \zeta_1^\kappa
\end{cases} \hspace{-10pt} , \quad \kappa=1,...,8\,,
\ee
where
\be \label{MF}
\quad M_F= \frac{M_B}{2\sqrt{2}} K(p^1) \, , \quad K(p^1) = \frac{4\pi\sqrt{2}M_B}{3 \, m_0 \, \beta} \frac{\widetilde H(p^1)}{\widetilde G(p^1)} = \frac{1+\frac{4\pi^2}{\beta^2}\alpha'^2 M^2 - \frac{4\pi}{\beta}\alpha' p^1}{\(1-\frac{2\pi}{\beta} \alpha' p^1\) \sqrt{1+\frac{4\pi^2}{\beta^2} \alpha'^2 M^2}} \, .
\ee
Plugging the first of \eqnref{eom} in the second one, and vice-versa, we can summarize them as
\be \label{kleingordondefcomponents}
\(-\eta^{\alpha\beta} \partial_\alpha \partial_\beta + M_F^2\) \zeta_I^\kappa = 0 \, , \quad I=1,2 \, , \quad \kappa=1,...,8 \, ,
\ee
from which
\be \label{kleingordondef}
\(-\eta^{\alpha\beta} \partial_\alpha \partial_\beta + M_F^2\) \psi = 0 \, , \quad \psi= \begin{pmatrix} 0 \\ \zeta_1 \\ 0 \\ \zeta_2 \end{pmatrix} \, .
\ee
The above equation can be interpreted as a free Klein-Gordon equation for a massive field $\psi$ of mass $M_F$. But how many propagating fermionic degrees of freedom does it describe? To sum up, we have introduced $\psi$ as a ten-dimensional Majorana spinor, so it is in principle composed by 32 real components. Then, we halve them imposing the gauge-fixing \eqnref{gaugefixing}. Finally, we can use the system of partial differential equations in \eqnref{eom} to constraint the general solution of \eqnref{kleingordondefcomponents}. So, if one knows the solution for $\zeta_1$, the one for $\zeta_2$ is automatically fixed and vice-versa. This halves further the degrees of freedom contained in $\psi$. We conclude that \eqnref{kleingordondef} describes eight fermionic degrees of freedom each with mass $M_F$. So, the number of fermionic propagating modes corresponds to the number of bosonic propagating modes, as required by supersymmetry.

Now, let us return to the mass-matching condition \eqnref{massmatchingcondition}. Notice that the fermionic mass in \eqnref{MF} is in general non-zero and depends on $p^1$, as well as on $\beta$. Nevertheless, making computations up to quadratic order in quantum fluctuations, we can not introduce further field dependence in \eqnref{ESF2}. So, the momentum $p^1$ appearing in the fermionic action is fixed by the classical version of the Virasoro constraints in \eqnref{classVirasoro}, that is\footnote{We will see how to generalize this relation to its quantum version. In particular, from \eqnref{quantumVirasoro}, it follows that $(p^1)^2 = \sum_{i=2}^d (p^i)^2 + \frac{\beta^2}{4\pi^2\alpha'^2} + o(\xi^2)$. Therefore, \eqnref{classp1} is exactly what we need at the current level of approximation.}
\be \label{classp1}
(p^1)^2 = \sum_{i=2}^d (p^i)^2 + \frac{\beta^2}{4\pi^2\alpha'^2} \, .
\ee
Since, to this order,
$K(p^1) = \sqrt{2}$,
we conclude that
\be
M_F = \frac{M_B}{2} \, .
\ee
Therefore, the mass-matching condition \eqnref{massmatchingcondition} is satisfied.\footnote{In \cite{PandoZayas:2003jr}, the mass-matching condition does not hold. This leads to somewhat different conclusions.}
This is a consistency check that the calculated masses are correct. 

\subsection{Summary}
\label{sec:summary}

Let us recapitulate the results of this section. We have concluded that
\begin{subequations} \label{ansatzdef}
  \begin{empheq}[left=\empheqlbrace]{align}
    &\label{X0} X^0 = m_0 \frac{\beta}{2\pi} \sigma \, , \quad X^0 \simeq X^0 + m_0\beta \, ,\\
    & X^i=\alpha' p^i m_0 \tau \, , \quad p^i \in \mathbb{R} \quad : \quad \widetilde \eta_{ij} \, p^i \, p^j = - M^2 \, , \quad i,j=1, ... , d \, ,\\
    &\label{U} U=u_0 \quad : \quad \left . g_{uu}(U) \, \partial_\alpha U \right |_{U=u_0} = c_\alpha \, , \quad \eta^{\alpha\beta} c_\alpha c_\beta = \frac{4}{9} \frac{1}{m_0^2} M_B^2 \, , \quad c_\tau \, , c_\sigma \in \mathbb{R} \, , \\
    &\label{ThetaDef} \Theta = \text{constant} \, ,
  \end{empheq}
\end{subequations}
is a classical configuration around which we can semi-classically quantize in a consistent way a closed string embedded in the Witten background \eqnref{Wback} and localized at the tip of the cigar. Indeed, one can expand the bosonic and fermionic action around it up to quadratic order in quantum fluctuations getting\footnote{\label{rescale2}Remember that the net effect of the diffeomorphism ghosts is to cancel out the contribution of two bosonic massless fluctuations. Moreover, we have rescaled the fermionic fluctuations as $\zeta_I^\kappa \mapsto \(m_0 \frac{\beta}{2\pi} \widetilde G (p^1)\)^{-1/2} \zeta_I^\kappa$.}
\begin{subequations}\label{finalS}
\begin{gather}
\begin{align}
&S_B^{(2)} = S_B^{cl} - \frac{1}{4 \pi \alpha'} \int  d\tau d\sigma \{ \eta^{\alpha\beta} \partial_\alpha \widetilde \xi^p \partial_\beta \widetilde \xi^q \delta_{pq} + M_B^2 \[(\widetilde \xi^u)^2 + (\widetilde \xi^\theta)^2\]  \} \, , \quad p, \, q=0,1,...,9 \, ,
\\
& \label{finalSF} S_F^{(2)}=\frac{i}{4\pi\alpha'} \int d\tau \, d\sigma \sum_{\kappa=1}^8 \{ -\zeta_1^\kappa ( - \partial_\tau + \partial_\sigma) \, \zeta_1^\kappa+\zeta_2^\kappa (\partial_\tau + \partial_\sigma) \, \zeta_2^\kappa + 2 \, (-1)^\kappa \, M_F \, \zeta_2^\kappa \zeta_1^\kappa \} \, , 
\end{align}
\end{gather}
\end{subequations}
where $S_B^{cl}$ is defined in \eqnref{SBcl} and
\be \label{finalmasses}
M_B^2 = \frac{9}{4} m_0^2 \frac{\beta^2}{4\pi^2} \, , \quad M_F= \frac{M_B}{2} \, .
\ee
The equations of motion for the (physical) quantum fluctuations are thus
\begin{subequations}
\begin{align}
&\label{finalbosoniceom}\(-\eta^{\alpha\beta} \partial_\alpha \partial_\beta + M_B^2\) \widetilde \xi^p = 0 \, , \quad p=u,\theta \, , \quad -\eta^{\alpha\beta} \partial_\alpha \partial_\beta \widetilde \xi^q = 0 \, , \quad q = 0,1,2,3,6,7 \, , \\
&\label{finalfermioniceom} \(-\eta^{\alpha\beta} \partial_\alpha \partial_\beta + M_F^2\) \zeta_I^\kappa = 0 \, , \,\,\, I=1,2 \, : 
\begin{cases}
\( - \partial_\tau + \, \partial_\sigma \) \zeta_1^\kappa  =   (-1)^{\kappa+1} \, M_F \, \zeta_2^\kappa\\
\( \partial_\tau + \, \partial_\sigma \) \zeta_2^\kappa  = (-1)^{\kappa+1} \, M_F \, \zeta_1^\kappa
\end{cases} \hspace{-10pt} , \,\,\, \kappa=1,...,8 \, .
\end{align}
\end{subequations}
So, the field content of these actions includes six massless bosons, two massive bosons of mass $M_B$ and eight massive fermions of mass $M_F$. As noted above, the spectrum displays the same number of bosonic and fermionic degrees of freedoms and satisfies the mass-matching condition \eqnref{massmatchingcondition}. For completeness, the conditions \eqnref{pbc} are trivial. Also the equations of motion \eqnref{classeom} for $k=0,1,...,d$ and $\theta$ are satisfied in a straightforward way in $U=u_0$, since $X^\mu$, $\mu=0,1,...,d$, is linear w.r.t.~$\tau$, $\sigma$ and $g_{\theta\theta}(u_0)=0$ (see \eqnref{eom1e2} and \eqnref{eom3}). Furthermore, the fall-off of $\partial_\alpha U$ in the tip limit and the condition for $c_\alpha$ specified in \eqnref{U} correspond to the equation of motion \eqnref{classeom} for $k=u$, that is \eqnref{eom4}. Let us stress again that the latter is formally valid regardless of whether quantum fluctuations are considered or not. 
See appendix \ref{appctau} for more details about $c_\tau$ and $c_\sigma$. Finally, the Virasoro constraints \eqnref{constraints} are satisfied $\forall M$ at classical level.

A final comment is in order.
The masses (\ref{finalmasses}) are essentially the same ones appearing in the quantum corrections to the string tension as derived from the Wilson loop in \cite{Bigazzi:2004ze}.
This is not a coincidence.
In fact, the classical winding configuration (\ref{ansatzdef}) that, as we have argued, should be the lighter one and so the one becoming tachyonic at smaller temperature, giving the value of the Hagedorn temperature $T_H$, is essentially the same as the one describing the Wilson loop.
As a consequence, one expects a direct link of $T_H$ to the field theory string tension $T_s$.
Although we have computed these observables in a specific case, their link is true in all the confining cases with a Type II gravity dual, since the only crucial ingredients are the existence of a Minkowskian part of the geometry and the fact that the string tension is non-vanishing in the IR (at the tip of the cigar in our case), which is the basic requirement for confinement. 

\section{Mass-shell condition and $T_H$} \label{secHagedorn}

Having fixed the classical reference configuration and its semi-classical quantization, we can dive in the computation of the Hagedorn temperature. As we have outlined in the introduction, the next step is trying to write a generalized mass-shell condition for closed strings localized at the tip of the cigar. Focusing only on the bosonic sector (for the moment), this can be achieved imposing the quantum version of \eqnref{Talphabeta0}. In particular, let us consider the combination
\be \label{T00T11}
\int d\tau \, d\sigma \(T^B_{\tau\tau} + T^B_{\sigma\sigma}\) = \int d\tau \, d\sigma \, \delta^{\alpha\beta} \partial_\alpha \rho^p \partial_\beta \rho^q g_{pq}(u) \, .
\ee
The above expression can be computed up to quadratic order in quantum fluctuations exploiting again the expansions in \eqnref{DuDuguudef}, \eqnref{DthetaDthetagthetatheta} and \eqnref{dxdxg00}. Nevertheless, we have to keep in mind that the indices $\alpha$ and $\beta$ are now implicitly contracted with $\delta^{\alpha\beta}$, instead of $\eta^{\alpha\beta}$. The result can be thus deduced by analogy from the tip limit of \eqnref{SB3}, that is (omitting the $o(\xi^2)$ terms for the sake of simplicity)
\be \label{quantumVirasoro}
\begin{split}
& \int \hspace{-4pt} d\tau d\sigma \(T^B_{\tau\tau} + T^B_{\sigma\sigma}\) = \\ & = \int \hspace{-4pt}  d\tau d\sigma \{ \delta^{\alpha\beta} \partial_\alpha X^\mu \partial_\beta X^\nu g_{\mu\nu}(u_0) + \delta^{\alpha\beta} \, \partial_\alpha \widetilde \xi^p \partial_\beta \widetilde \xi^q \, \delta_{pq} + \frac{9}{4} \, m_0^2 \, \delta^{\alpha\beta} c_\alpha c_\beta \[(\widetilde \xi^u)^2+(\widetilde \xi^\theta)^2\] \} \, ,
\end{split}
\ee
where $\mu$, $\nu=0,1,...,d$, $p$, $q=0,1,...,9$ and, from \eqnref{ansatzdef},
\be \label{orderzero}
\delta^{\alpha\beta} \partial_\alpha X^\mu \partial_\beta X^\nu g_{\mu\nu}(u_0) = \[\frac{\beta^2}{4\pi^2}-\alpha'^2 M^2\] \( \frac{u_0}{R} \)^{\sfrac{3}{2}}
\ee
and
\be
\frac{9}{4} \, m_0^2 \, \delta^{\alpha\beta} c_\alpha c_\beta = M_B^2 + \frac{9}{2} \, m_0^2 \, c_\tau^2 \, .
\ee

The on-shell bosonic fluctuations are given by the solutions of \eqnref{finalbosoniceom}, that is
\begin{subequations} \label{solKleinGordon}
\begin{gather}
\begin{align}
& \widetilde \xi^p = \widetilde \xi^p_0 + \widetilde \Xi^p \, , && \hspace{-280pt} p=u,\theta \, , \quad \omega_n = \text{sgn}(n) \sqrt{n^2+M_B^2} \, , \nb \\ 
& \, && \hspace{-280pt} \widetilde \xi^p_0 = i \sqrt{\frac{\alpha'}{2}} \sqrt{\frac{1}{M_B}} \( a^p e^{- i \, M_B \, \tau} - {a^p}^\dagger e^{+ i \, M_B \, \tau} \) , \nb \\
& \, && \hspace{-280pt} \widetilde \Xi^p = i \, \sqrt{\frac{\alpha'}{2}} \sum_{n \ne 0} \frac{1}{\omega_n} \(\alpha_n^p \, e^{-i \, n \, \sigma} + \widetilde \alpha_n^p \, e^{+i \, n \, \sigma}\) e^{- i \, \omega_n \, \tau} \, , \\
&\widetilde \xi^q = i \, \sqrt{\frac{\alpha'}{2}} \sum_{n \ne 0} \frac{1}{n} \(\alpha_n^q \, e^{-i \, n \, \sigma} + \widetilde \alpha_n^q \, e^{+i \, n \, \sigma}\) e^{- i \, n \, \tau} \, , \quad q=0,1,2,3,6,7  \, ,
\end{align}
\end{gather}
\end{subequations}
where
\begin{subequations}\label{brackets}
\begin{gather}
\begin{align}
& \[\alpha_n^p, \alpha_m^p\]=\[\widetilde \alpha_n^p, \widetilde \alpha_m^p\] = \delta_{m+n,0} \, \omega_n \, , \quad \[a^p, {a^p}^\dagger\]=1 \, \quad p=u,\theta \, ,\\
& \[\alpha_n^q, \alpha_m^q\]=\[\widetilde \alpha_n^q, \widetilde \alpha_m^q\] = \delta_{m+n,0} \, n \, , \quad q=0,1,2,3,6,7 \, .
\end{align}
\end{gather}
\end{subequations}
Given these conventions, the bosonic quantum fluctuations respect the canonical commutation relations
\be \label{cancommrel}
\[\widetilde \xi^l (\tau,\sigma), \partial_\tau \widetilde \xi^{l'} (\tau,\sigma')\] = 2 \pi \alpha' i \delta^{ll'} \delta\(\sigma-\sigma'\)\, , \quad \forall l , l' \, .
\ee

Let us notice that
\be
\int_0^{2\pi} \hspace{-8pt} d\sigma \, \widetilde \xi^p_0 \, \widetilde \Xi^p \sim \sum_{n \ne 0} \int_0^{2\pi} \hspace{-8pt} d\sigma e^{\pm i \, n \, \sigma} = 2 \pi \sum_{n \ne 0} \delta(n) = 0 \, .
\ee
So, what we have to compute is
\be
\int_0^{2\pi} \hspace{-8pt} d\sigma \{\(\partial_\tau \widetilde \xi^p_0\)^2 +\[M_B^2 + \frac{9}{2} \, m_0^2 \, c_\tau^2\] \(\widetilde \xi^p_0\)^2 \} = 2 \pi \alpha' \[\(1+\frac{9}{4}\frac{m_0^2 \, c_\tau^2}{M_B^2}\) A^p_0 - \frac{9}{4}\frac{m_0^2 \, c_\tau^2}{M_B^2} B^p_0\] \, ,
\ee
for $p=u$, $\theta$ and
\be
A_0^p = M_B \( a^p \, {a^p}^\dagger + {a^p}^\dagger \, a^p \) \, , \quad B_0^p = M_B \(a^p \, a^p e^{-2 \, i \, M_B \, \tau} + {a^p}^\dagger \, {a^p}^\dagger e^{+2 \, i \, M_B \, \tau}\) \, .
\ee
Moreover,\footnote{We have used that $\omega_{-n}=-\omega_n$ and $\int_0^{2\pi} \hspace{-6pt} d\sigma e^{- i \, (n+m) \, \tau} = 2\pi\delta_{n+m,0}$.}
\be \label{intdsigmaxi2}
\begin{split}
& \int_0^{2\pi} \hspace{-8pt} d\sigma \(\widetilde \Xi^p\)^2 
= \int_0^{2\pi} \hspace{-8pt} d\sigma \hspace{-2pt} \sum_{n, m \ne 0} \frac{-\sfrac{\alpha'}{2}}{\omega_n \omega_m} \(\alpha_n^p \, e^{-i \, n \, \sigma} + \widetilde \alpha_n^p \, e^{+i \, n \, \sigma}\) \(\alpha_m^p \, e^{-i \, m \, \sigma} + \widetilde \alpha_m^p \, e^{+i \, m \, \sigma}\) e^{- i \, (\omega_n+\omega_m) \, \tau} \\
&= \int_0^{2\pi} \hspace{-8pt} d\sigma \hspace{-2pt} \sum_{n, m \ne 0} \frac{-\sfrac{\alpha'}{2}}{\omega_n \omega_m} \(\alpha_n^p \, e^{-i \, \omega_n \, \tau} - \widetilde \alpha_{-n}^p \, e^{+i \, \omega_n \, \tau}\) \(\alpha_m^p \, e^{-i \, \omega_m \, \tau} - \widetilde \alpha_{-m}^p \, e^{+i \, \omega_m \, \tau}\) e^{- i \, (n+m) \, \sigma}\\
&= \pi\alpha' \sum_{n \ne 0} \frac{1}{\omega_n^2} \(\alpha_n^p \, e^{-i \, \omega_n \, \tau} - \widetilde \alpha_{-n}^p \, e^{+i \, \omega_n \, \tau}\) \(\alpha_{-n}^p \, e^{+i \, \omega_n \, \tau} - \widetilde \alpha_{n}^p \, e^{-i \, \omega_n \, \tau}\)\\
&= \pi\alpha' \sum_{n \ne 0} \frac{1}{\omega_n^2} \( A_n^p - B_n^p \)
\end{split} \, ,
\ee
for $p=u,\theta$ and
\be
A_n^p = \alpha_n^p \, \alpha_{-n}^p + \widetilde \alpha_{-n}^p \, \widetilde \alpha_n^p \, , \quad B_n^p = \alpha_n^p \, \widetilde \alpha_n^p e^{-2 \, i \, \omega_n \, \tau} + \widetilde \alpha_{-n}^p \, \alpha_{-n}^p e^{+2 \, i \, \omega_n \, \tau} \, .
\ee
Similarly, we get
\be
\int_0^{2\pi} \hspace{-8pt} d\sigma \( \partial_\tau \widetilde \Xi^p\)^2 = \pi\alpha' \sum_{n \ne 0} \( A_n^p + B_n^p \) \, , \quad \int_0^{2\pi} \hspace{-8pt} d\sigma \( \partial_\sigma \widetilde \Xi^p\)^2 = \pi\alpha' \sum_{n \ne 0} \frac{n^2}{\omega_n^2} \( A_n^p - B_n^p \)  \, , \quad p=u,\theta \, .
\ee
The results for $q=0,1,2,3,6,7$ can be deduced from the above ones taking the $M_B \to 0$ limit. So we obtain
\be
\hspace{-4pt} \int \hspace{-4pt} d\sigma \hspace{-4pt} \[\sum_{\substack{p=u,\theta \\ \alpha=\tau,\sigma}} \hspace{-4pt} \partial_\alpha \widetilde \Xi^p \partial_\alpha \widetilde \Xi^p \hspace{-4pt} + \hspace{-20pt} \sum_{\substack{q=0,1,2,3,6,7 \\ \alpha = \tau, \sigma}} \hspace{-16pt} \partial_\alpha \widetilde \xi^q \partial_\alpha \widetilde \xi^q\] \hspace{-4pt}= \pi \alpha' \hspace{-2pt} \sum_{n \ne 0} \{\sum_{p=u,\theta} \[\[1+\frac{n^2}{\omega_n^2}\] \hspace{-2pt} A_n^p + \hspace{-2pt} \[1-\frac{n^2}{\omega_n^2}\] \hspace{-2pt} B_n^p \] + \hspace{-16pt} \sum_{q=0,1,2,3,6,7} \hspace{-16pt} 2 A_n^q\} .
\ee
Finally, let us note that the holographic dictionary states that
\be \label{dictionary}
\( \frac{u_0}{R} \)^{\sfrac{3}{2}} = \alpha' \frac{\lambda}{6} m_0^2 \, , \quad \lambda = 4 \pi g_s M_{KK} \sqrt{\alpha'} N \, ,
\ee
where $\lambda$ is  the 't Hooft coupling of the WYM theory. All in all,
\be \label{T00T11expanded}
\begin{split}
&\frac{1}{2\pi\alpha'} \int d\tau \, d\sigma \(T^B_{\tau\tau} + T^B_{\sigma\sigma}\) = \\
&= \hspace{-4pt} \int \hspace{-2pt} d\tau \{ \sum_{n \in \mathbb{Z}}\[\sum_{p=u,\theta} \[\(1+\frac{9}{4}\frac{m_0^2 c_\tau^2}{\omega_n^2}\) A_n^p - \frac{9}{4}\frac{m_0^2 c_\tau^2}{\omega_n^2} B_n^p\] + \hspace{-4pt} \sum_{q=0,1,2,3,6,7} \hspace{-10pt} A_n^q   \] + \frac{\lambda}{6} m_0^2 \[\frac{\beta^2}{4\pi^2}-\alpha'^2 M^2\]\}
\end{split}
 \, .
\ee
Notice that so far we have met no constraints on $c_\tau$ alone. Nevertheless, in appendix \ref{appctau} we see that 
\be
c_\tau = 0
\ee
is a necessary condition in order not to have tachyons in the spectrum. En passant, thanks to \eqnref{U}, this fixes also
\be
c_\sigma = \pm \frac{2}{3} \frac{1}{m_0} M_B = \pm \frac{M_B}{M_{KK}} \, .
\ee
Basically, we are dealing with the Hamiltonian $H_B$ which describes the dynamics of the bosonic excitations in a closed string theory, since
\be
H_B = \frac{1}{4\pi\alpha'} \int d\sigma \(T^B_{\tau\tau} + T^B_{\sigma\sigma}\) \, .
\ee
This choice puts the Hamiltonian in a standard \emph{diagonal} form and we can rewrite \eqnref{T00T11expanded} as
\be
\int d\tau \, H_B = \frac{1}{2} \int \hspace{-4pt} d\tau \{\sum_{n \in \mathbb{Z}} \[\sum_{p=u,\theta} A_n^p + \hspace{-4pt} \sum_{q=0,1,2,3,6,7} \hspace{-10pt} A_n^q   \] + \frac{\lambda}{6} m_0^2 \[\frac{\beta^2}{4\pi^2}-\alpha'^2 M^2\]\}
 \, .
\ee
Since the integrand does not depend on $\tau$, we have
\be \label{T00T11diagonal}
H_B =\frac{1}{2} \sum_{n \in \mathbb{Z}} \[\sum_{p=u,\theta} A_n^p + \hspace{-4pt} \sum_{q=0,1,2,3,6,7} \hspace{-10pt} A_n^q   \] + \frac{\lambda}{12} m_0^2 \[\frac{\beta^2}{4\pi^2}-\alpha'^2 M^2\]
 \, .
\ee

At this point, we just have to manipulate
\be
A_0^p = M_B \(2 {a^p}^\dagger a^p + [a^p, {a^p}^\dagger]\) \, , \quad p=u,\theta\,,
\ee
and
\be
\begin{split}
\sum_{n \ne 0} A_n^r 
&= \sum_{n > 0} \(\alpha_n^r \, \alpha_{-n}^r + \widetilde \alpha_{-n}^r \, \widetilde \alpha_n^r\) + \sum_{n < 0} \(\alpha_n^r \, \alpha_{-n}^r + \widetilde \alpha_{-n}^r \, \widetilde \alpha_n^r\) \\
&= \sum_{n > 0} \(\alpha_n^r \, \alpha_{-n}^r + \alpha_{-n}^r \, \alpha_n^r + \widetilde \alpha_{-n}^r \, \widetilde \alpha_n^r + \widetilde \alpha_n^r \, \widetilde \alpha_{-n}^r\)\\
&= \sum_{n > 0} \(\[\alpha_n^r , \alpha_{-n}^r\] + 2 \, \alpha_{-n}^r \, \alpha_n^r + \[\widetilde \alpha_n^r , \widetilde \alpha_{-n}^r\] + 2 \, \widetilde \alpha_{-n}^r \, \widetilde \alpha_n^r\)\\
\end{split} \, .
\ee
Exploiting \eqnref{brackets}, we conclude that
\be \label{finalH}
H_B = N_B^{(0)} + N_B + \widetilde N_B + \frac{\lambda}{12} m_0^2 \[\frac{\beta^2}{4\pi^2}+\alpha'^2 \widetilde \eta_{ij} p^i p^j\]+ \, \sum_{n \in \mathbb{Z}} |\omega_n| + 3 \, \sum_{n \in \mathbb{Z}} |n|
 \, ,
\ee
where
\begin{subequations}
\begin{gather}
\begin{align}
&N_B^{(0)} = M_B \sum_{p=u,\theta} {a^p}^\dagger a^p \, , \\
& N_B=\sum_{n = 1}^{+\infty} \[\sum_{p=u,\theta} \alpha_{-n}^p \alpha_n^p + \hspace{-4pt} \sum_{q=0,1,2,3,6,7} \hspace{-10pt} \alpha_{-n}^q \alpha_n^q\] \, ,\\
& \widetilde N_B=\sum_{n = 1}^{+\infty} \[\sum_{p=u,\theta} \widetilde \alpha_{-n}^p \widetilde \alpha_n^p + \hspace{-4pt} \sum_{q=0,1,2,3,6,7} \hspace{-10pt} \widetilde \alpha_{-n}^q \widetilde \alpha_n^q\] \, .
\end{align}
\end{gather}
\end{subequations}

Now we should compute the fermionic contribution to the total string stress-energy tensor and add it to the bosonic one. Nevertheless, let us infer the result quickly. From \eqnref{finalSF}, we define the Lagrangian of the fermionic sector as
\be
L_F = \frac{i}{4\pi\alpha'} \int d\sigma \sum_{\kappa=1}^8 \{ -\zeta_1^\kappa ( - \partial_\tau + \partial_\sigma) \, \zeta_1^\kappa+\zeta_2^\kappa (\partial_\tau + \partial_\sigma) \, \zeta_2^\kappa + 2 \, (-1)^\kappa \, M_F \, \zeta_2^\kappa \zeta_1^\kappa \} \, .
\ee
Given \eqnref{finalfermioniceom}, it is straightforward to see that the on-shell fermionic Lagrangian is zero. Therefore, with
\be \label{canfermmom}
\Pi_I^\kappa = - \frac{\partial \, \mathcal{L}_F}{\partial \(\partial_\tau \zeta_I^\kappa\)} = + \frac{i}{4\pi\alpha'} \, \zeta_I^\kappa \, ,
\ee
the on-shell fermionic Hamiltonian simply reads
\be \label{canonicalHF}
H_F = \int d\sigma \sum_{\kappa=1}^8 \sum_{I=1}^2 \Pi_I^\kappa \, \partial_\tau \zeta_I^\kappa = + \frac{i}{4\pi\alpha'} \int d\sigma \sum_{\kappa=1}^8 \sum_{I=1}^2 \zeta_I^\kappa \, \partial_\tau \zeta_I^\kappa \, .
\ee
So, we have just to specify the mode expansion of $\zeta_I^\kappa$. Let us stress that, in field theory, spacetime fermions at finite temperature are taken with anti-periodic boundary conditions along the thermal circle.\footnote{For example, see Appendix A of \cite{Dashen:1975xh} for a nice argument. Atick and Witten emphasize this point in \cite{Atick:1988si}. In particular, working in the RNS formalism, they deform the standard GSO projections in order to preserve and implement this principle in string theory.} In our conventions, this means that $\zeta^k_I$ is anti-periodic under $X^0 \mapsto X^0 + m_0 \, \beta$ for all $I=1, 2$, $k=1,...,8$. Given the ansatz for $X^0$ in \eqnref{X0}, the anti-periodic boundary condition corresponds to
\be \label{abc}
\zeta_I^k(\tau,\sigma+2\pi)=-\zeta_I^\kappa(\tau,\sigma) \, , \quad \forall \, I=1,2 \, , \, \kappa=1,..,8 \, .
\ee
A solution of \eqnref{finalfermioniceom} compatible with \eqnref{abc} is
\begin{subequations} \label{solFermions}
  \begin{empheq}[left=\empheqlbrace]{align}
    &\zeta_1^\kappa = \sum_{\substack{r \in \mathbb{Z}+\frac{1}{2}}} c_r \[ S_r^\kappa \, e^{-i \, r \, \sigma} + \frac{i}{M_F} (\omega_r-r) \, \widetilde S_r^\kappa \, e^{+i \, r \, \sigma} \] e^{-i \, \omega_r \, \tau} \, , \\
    &\zeta_2^\kappa = (-1)^{\kappa} \sum_{\substack{r \in \mathbb{Z}+\frac{1}{2}}} c_r \[ \widetilde S_r^\kappa \, e^{+i \, r \, \sigma} - \frac{i}{M_F} (\omega_r - r) \, S_r^\kappa \, e^{-i \, r \, \sigma} \] e^{-i \, \omega_r \, \tau} \, ,
  \end{empheq}
\end{subequations}
where $\kappa=1,...8$,
\be
\omega_r=\text{sgn}(r)\sqrt{r^2+ M_F^2} \, , \quad c_r = \sqrt{\frac{\alpha' M_F^2}{M_F^2 + (\omega_r-r)^2}}
\ee
and
\be
\{S_r^\kappa, S_s^{\kappa'}\} = \{\widetilde S_r^\kappa, \widetilde S_s^{\kappa'}\} = \delta^{\kappa\kappa'} \delta_{r+s,0} \, .
\ee
With this conventions
\be
\{\zeta_I^\kappa(\tau,\sigma),\zeta_J^{\kappa'}(\tau,\sigma')\}=2\pi\alpha' \delta^{\kappa\kappa'} \delta_{IJ} \delta(\sigma-\sigma') \, .
\ee
So, let us start computing
{\small{
\be \label{zeta1Dtauzeta1}
\begin{split}
&\int_0^{2\pi} \hspace{-10pt} d\sigma \zeta_1^\kappa \partial_\tau \zeta_1^\kappa =\\
&= 2 \pi i \hspace{-4pt} \sum_{r \in \mathbb{Z}+\frac{1}{2}} \hspace{-6pt} c_r^2 \, \omega_r \[ S_r^\kappa S_{-r}^\kappa - \frac{1}{M_F^2} (\omega_r-r)^2 \widetilde S_{-r}^\kappa \widetilde S_r^\kappa  - \frac{i}{M_F} (\omega_r-r) (S_r^\kappa \widetilde S_r^\kappa e^{-2 i \omega_r \tau} + \widetilde S_{-r}^\kappa S_{-r}^\kappa e^{+2 i \omega_r \tau}) \]
\end{split} \, .
\ee
}}Similarly,
{\small{
\be
\begin{split} \label{zeta2Dtauzeta2}
& \int_0^{2\pi} \hspace{-10pt} d\sigma \zeta_2^\kappa \partial_\tau \zeta_2^\kappa =  \\ 
& = 2 \pi i \hspace{-4pt} \sum_{r \in \mathbb{Z}+\frac{1}{2}} \hspace{-6pt} c_r^2 \, \omega_r \[ -  \widetilde S_{-r}^\kappa \widetilde S_r^\kappa + \frac{1}{M_F^2} (\omega_r-r)^2 S_r^\kappa S_{-r}^\kappa  + \frac{i}{M_F} (\omega_r-r) (S_r^\kappa \widetilde S_r^\kappa e^{-2 i \omega_r \tau} + \widetilde S_{-r}^\kappa S_{-r}^\kappa e^{+2 i \omega_r \tau}) \]\end{split} \, . 
\ee
}}Therefore,
\be \label{HFprime}
\begin{split}
H_F 
&=  + \frac{i}{4\pi\alpha'} \sum_{\kappa=1}^8 2\pi i \hspace{-4pt} \sum_{r \in \mathbb{Z}+\frac{1}{2}} c_r^2 \, \omega_r \(1+\frac{1}{M_F^2}(\omega_r-r)^2\) \[S_r^\kappa S_{-r}^\kappa - \widetilde S_{-r}^\kappa \widetilde S_r^\kappa\] \\
&= -\frac{1}{2} \sum_{r \in \mathbb{Z}+\frac{1}{2}} \omega_r \[S_r^\kappa S_{-r}^\kappa - \widetilde S_{-r}^\kappa \widetilde S_r^\kappa\] \\
&= \sum_{r = \frac{1}{2}}^{+\infty} \omega_r \[S_{-r}^\kappa S_r^\kappa  + \widetilde S_{-r}^\kappa \widetilde S_r^\kappa - \{S_r^\kappa, S_{-r}^\kappa\}\] \, .
\end{split}
\ee
We conclude that
\be
H_F = N_F + \widetilde N_F - 4 \sum_{r \in \mathbb{Z}+\frac{1}{2}} |\omega_r| \, ,
\ee
where
\begin{subequations}
\begin{gather}
\begin{align}
& N_F=\sum_{r = \frac{1}{2}}^{+\infty} \omega_r \, S_{-r}^\kappa S_r^\kappa \, ,\\
& \widetilde N_F=\sum_{r = \frac{1}{2}}^{+\infty} \omega_r \, \widetilde S_{-r}^\kappa \widetilde S_r^\kappa \, .
\end{align}
\end{gather}
\end{subequations}

All in all, the total Hamiltonian $H=H_B + H_F$ is thus
\be \label{totalH}
H=N_B^{(0)} + N_B + \widetilde N_B + N_F + \widetilde N_F + \frac{\lambda}{12} m_0^2 \[\frac{\beta^2}{4\pi^2}+\alpha'^2 \widetilde \eta_{ij} p^i p^j\]+ \, \sum_{n \in \mathbb{Z}} |\omega_n| + 3 \, \sum_{n \in \mathbb{Z}} |n| - 4 \sum_{r \in \mathbb{Z}+\frac{1}{2}} |\omega_r| \, .
\ee
Now, let us introduce
{\small{
\be \label{Delta}
\Delta_a \( \mathcal{M} \) = \frac{1}{2} \sum_{n \in \mathbb{Z}} \sqrt{\mathcal{M}^2+\(n+a\)^2} - \frac{1}{2} \int_{-\infty}^{+\infty} \hspace{-10pt} dk \, \sqrt{\mathcal{M}^2+(k+a)^2} = - \frac{\mathcal{M}}{\pi} \, \sum_{p=1}^{+\infty} \, \frac{\text{cos}(2 \pi \, a \, p)}{p} \, K_1(2 \pi \, \mathcal{M} \, p) \, ,
\ee
}}as the regularized form of the zero point (or Casimir) energy for two-dimensional massive boson ($a=0$) or fermion ($a=1/2$) fields. Therefore, the regularization procedure realizes in
\be
\sum_{n \in \mathbb{Z}} |\omega_n| \mapsto 2 \, \Delta_0(M_B) \, , \quad \sum_{n \in \mathbb{Z}} |n| \mapsto 2 \, \Delta_0(0) \, , \quad \sum_{r \in \mathbb{Z}+\frac{1}{2}} |\omega_r| \mapsto 2 \, \Delta_{\sfrac{1}{2}}(M_F) \, .
\ee
More explicitly, \eqnref{Delta} can be expressed in terms of a power series as \cite{Hyun:2003ks}
\be \label{Casimirpowerseries}
{\tiny{\Delta_a \hspace{-2pt} \( \mathcal{M} \) =}}
\begin{cases}
\hspace{-4pt} - \frac{1}{12} + \frac{\mathcal{M}}{2} + \frac{\mathcal{M}^2}{4} \[ \log \frac{\mathcal{M}^2}{4} + 2 \, \gamma_E - 1 \] \hspace{-4pt} \, + \, \hspace{-4pt} \underset{k=2}{\overset{+\infty}{\sum}} B_k \, \zeta( 2k-1 ) \mathcal{M}^{2k} &\hspace{-10pt} (a=0) \\ 
\hspace{-4pt} \frac{1}{24} + \frac{\mathcal{M}^2}{4} {\tiny{\[ \log \frac{\mathcal{M}^2}{4} +2 \, \gamma_E - 1 + 4 \, \log2 \] \hspace{-4pt} \, + \, \hspace{-4pt} \underset{k=2}{\overset{+\infty}{\sum}} B_k \(2^{2k-1}-1\) \zeta( 2k-1 ) \mathcal{M}^{2k}}} &\hspace{-10pt}(a=\frac{1}{2})
\end{cases},
\ee
where $\gamma_E$ is the Euler constant, $\zeta$ is the Riemann zeta function and $B_k =\frac{(-1)^k \Gamma\(k-\frac{1}{2}\)}{\Gamma \(-\frac{1}{2}\) \Gamma \( k+1 \)}$.

In this way, the constraint $H=0$ reads
\be \label{massshellcond}
N_B^{(0)} + N_B + \widetilde N_B + N_F + \widetilde N_F +\frac{\lambda}{12} m_0^2 \[\frac{\beta^2}{4\pi^2}-\alpha'^2 M^2\]+ 2 \, \Delta_0 \( M_B \) + 6 \, \Delta_0 \( 0 \) - 8 \Delta_{\sfrac{1}{2}} \( M_F \)= 0
 \, ,
\ee
and represents the desired generalized mass-shell condition for closed strings embedded in the finite temperature Witten background at the tip of the cigar. Basically, it is the quantum version of \eqnref{classVirasoro}. Let us stress that \eqnref{massshellcond} holds $\forall M^2$.

\subsection{The Hagedorn temperature}

Since we are looking for the lighter state going massless, we focus on the $N_B^{(0)}=N_B=\widetilde N_B=N_F=\widetilde N_F = 0$ level. Taking the $M^2 \to 0$ limit and considering that $\beta_H^{class}=0$,
the Hagedorn temperature of the WYM theory, to first order in the worldsheet quantum corrections, is thus given by
\be \label{TH}
\frac{\lambda}{12} m_0^2 \frac{\beta_H^2}{4\pi^2} + 2 \, \Delta_0 \( 0 \) + 6 \, \Delta_0 \( 0 \) - 8 \, \Delta_{\sfrac{1}{2}} \( 0 \) = 0
\, .
\ee

This formula can be checked in the \emph{flat limit} (i.e. superstring theory in flat space taking into account quantum corrections). Exploiting \eqnref{dictionary}, the latter corresponds to
\be \label{flatlimit}
\alpha' \frac{\lambda}{6} m_0^2=m_0^2 \, g_{00}(u_0) \to 1 \, .
\ee
In this regime, \eqnref{TH} reduces to
\be
\frac{\(\beta_H^{flat}\)^2}{8\pi^2\alpha'} =- 8 \( \Delta_0 \( 0 \) - \Delta_{\sfrac{1}{2}} \( 0 \)\) = -8 \(-\frac{1}{12}-\frac{1}{24}\) = 1  \, ,
\ee
from which
\be \label{THflat}
T_H^{flat}=\frac{1}{\pi\sqrt{8\alpha'}} = \sqrt{\frac{T_s}{4\pi}} \, , \quad T_s = \frac{1}{2\pi\alpha'} \, .
\ee
This is exactly the outcome of flat space reported in \cite{Sundborg:1984uk,Bowick:1985az,Tye:1985jv,Matsuo:1986es,Sathiapalan:1986db,Kogan:1987jd,OBrien:1987kzw,Atick:1988si}. In particular, it is the same result found by Atick and Witten in the context of Type II RNS superstring theory and deforming the GSO projections in order to include periodic bosons and anti-periodic fermions in the spectrum.

In the general case, the final result from (\ref{TH}) is
\be \label{THeff}
\boxed{
T_H = \sqrt{\frac{T_s}{4\pi}} 
\quad \Rightarrow \quad \overline T_H = \frac{T_H}{T_c} = \sqrt{\frac{\lambda}{27}}
}
\, ,
\ee
where
\be \label{effTs}
T_s = \frac{m_0^2 \, g_{00}(u_0)}{2\pi\alpha'} = \frac{m_0^2 \lambda}{12\pi}
\ee
is the gauge theory string tension in the WYM theory\footnote{Given \eqnref{flatlimit}, the effective string tension \eqnref{effTs} goes to $1/2\pi\alpha'$ in the flat limit.} 
and 
\be
T_c=\frac{M_{KK}}{2\pi}
\ee 
is the critical temperature for deconfinement.
Let us observe that \eqnref{THeff} corresponds to \eqnref{THflat} once the flat space string tension $T_s=1/2\pi\alpha'$ has been replaced with the effective one defined in \eqnref{effTs}. 

\begin{figure}[t]
\centering
{\includegraphics[width=0.6\textwidth]{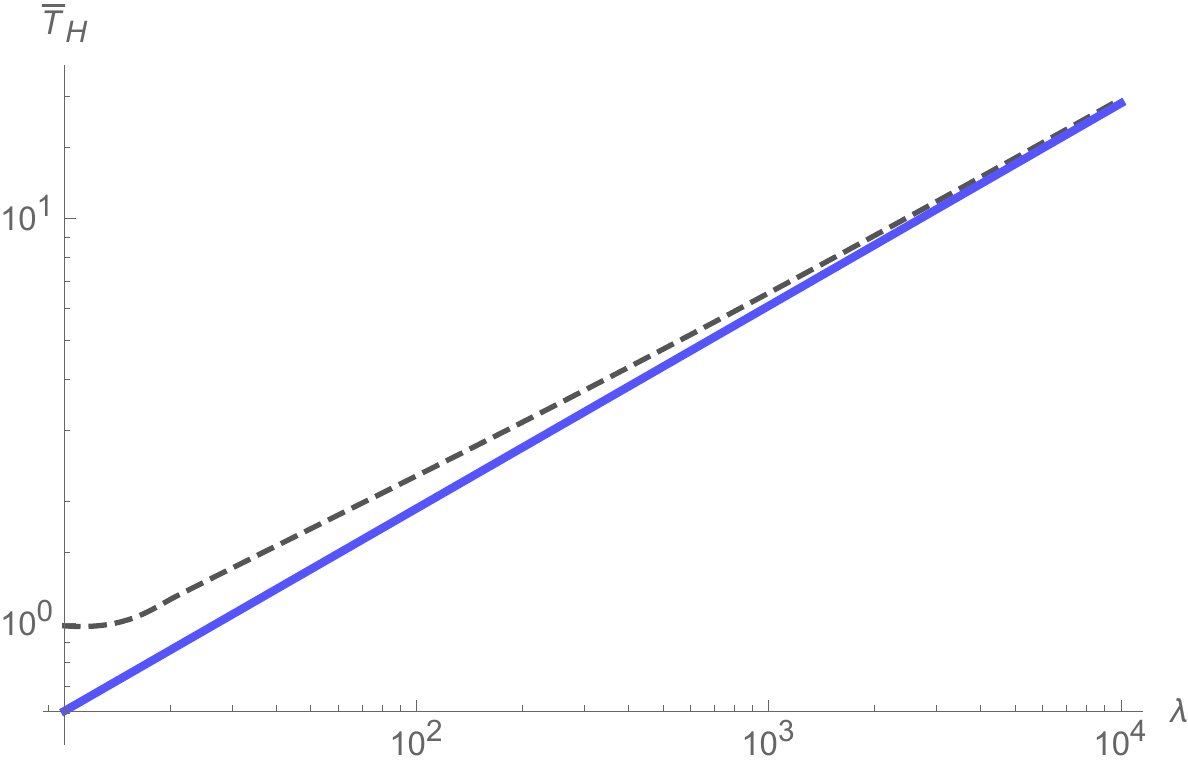}\label{THvsTHflat}}
\caption{Log-log plot of the Hagedorn temperature $\overline T_H$ from \eqnref{THeff} (solid blue line) as the `t Hooft coupling $\lambda$ varies. The dashed line reports the result for $\overline T_H$ in \eqnref{THexpanded}, which includes a subset of $1/\lambda$ corrections. These results are obtained keeping only the three hundred leading terms in the expansion \eqnref{THexpanded}.} 
\label{plotTH}
\end{figure}

Note that in deriving (\ref{TH}) from (\ref{massshellcond}) we have discarded $\mathcal{O}(\lambda^{-1/2})$ corrections for consistency, since our quadratic-order calculation does not automatically include all of these contributions.
Nevertheless, $M_B$ encodes a subset of these corrections, which can be included in $T_H$ via  
\be
\frac{\lambda}{12} m_0^2 \frac{\beta_H^2}{4\pi^2} + 2 \, \Delta_0 \( \frac{3}{2} m_0 \frac{\beta_H}{2\pi} \) + 6 \, \Delta_0 \( 0 \) - 8 \, \Delta_{\sfrac{1}{2}} \( \frac{3}{4} m_0 \frac{\beta_H}{2\pi} \) = 0
\, .
\ee
In order to get a suitable expression for numerical computations, we can plug \eqnref{Casimirpowerseries} into the above expression getting\footnote{Notice that \eqnref{THexpanded} has the same structure of the results one can find in the pp-wave literature. The thermodynamics of Type IIA GS superstring on the ten dimensional pp-wave geometry has been studied in \cite{Hyun:2003ks,Hyun:2002wu,Shin:2003ae}. Similar results have been also found in the Type IIB case in \cite{PandoZayas:2002hh, Greene:2002cd, Bigazzi:2003jk}. }
\be \label{THexpanded}
\frac{\lambda}{54 \, \overline T_H^2} - \frac{1}{2} \( 1 - \frac{1}{\overline T_H} +  \frac{\log2}{\overline T_H^2} \) + \underset{k=2}{\overset{+\infty}{\sum}} B_k \(4^{1-k}-1\) \zeta( 2k-1 ) \frac{1}{\overline T_H^{2k}} = 0
\, .
\ee
The comparison between \eqnref{THeff} and the outcome of \eqnref{THexpanded} is resumed 
in figure \ref{plotTH}. The corrections due to the mass of the world-sheet modes are obviously important for not too large values of $\lambda$ and increase the value of $T_H$. In the $\lambda \to + \infty$ limit the two results coincide. 

\section{Conclusions}
\label{secConclusions}

In general, we do not know how to quantize exactly a string embedded in a curved space. This means that every genus-one computation of $T_H$ is in principle precluded and we can not proceed as usual in flat space or e.g.~in pp-wave backgrounds. Nevertheless, Atick and Witten proposed in \cite{Atick:1988si} an alternative genus-zero approach which can be extended to superstring theories in curved space. Employing and generalizing this method, we have provided an estimate of the Hagedorn temperature for the WYM theory making computations in the dual Type IIA superstring theory on the Witten background, at leading order in quantum corrections. 

This derivation requires the knowledge of the mass-shell condition of the theory. In this way, following the winding interpretation of the problem, $T_H$ has been deduced in \eqnref{THeff} fixing to zero the mass of the physical ground state. Notice that the mass-shell condition in \eqnref{massshellcond} has been derived given the field content of the string action expanded around a reference classical configuration up to quadratic order in quantum fluctuations, as in \eqnref{finalS}. The classical configuration has been found in section \ref{sec:trulyclassconf} and resumed in section \ref{sec:summary}. Its final version is given in appendix \ref{appctau}.

Despite the limits of this computation, let us stress again that our proposal is in agreement with the widespread expectation in the literature and fixes its ambiguities in an analytical way. Moreover, it crucially depends only on the presence of a Minkowskian sector in the background metric and on the non-vanishing limit of the string tension in the low energy regime. These proprieties are shared with every confining theory having a Type II holographic counterpart. So, at leading order in $\alpha'$ and supported by a consistent full computation, we claim that the Hagedorn temperature of whatever Type II stringy model dual to a confining theory can be estimated as \eqnref{result}, where the flat space string tension has been replaced with the effective one in curved space.

In the future, it would be interesting to estimate the density of states in the theories we have considered.
Knowledge of the Hagedorn temperature and the density of states would allow, among other applications, to evaluate the reheating temperature in very supercooled confining transitions.
In line of principles, the latter could source detectable gravitational waves, in the case that dark matter has a component described by a confining gauge theory (e.g., see \cite{Forestell2017rdg}). 
Indeed, this scenario, which we have started exploring in \cite{Bigazzi:2020phm,Bigazzi:2020avc,Bigazzi:2021bwv} within a top-down holographic context, has been part of the motivation for the present work.

\vskip 15pt \centerline{\bf Acknowledgments} \vskip 10pt 

\noindent 
We thank Domenico Seminara, Luca Martucci and especially Valentina Giangreco Puletti and Wolfgang Mueck for comments and very helpful discussions.


\appendix
\section{About the spin-connection and the Ramond-Ramond field strength of the Witten background}
\label{appA}
\subsection{The spin-connection} \label{appspinconn}
In this section we aim to compute every component of the spin-connection $\omega_{\underline a \underline b, \, p}$ as a function of the components of the Witten background metric taken with Lorentzian signature \eqnref{Wback} and given the approximation \eqnref{Omegaapprox}. In general, the following computation for $\omega_{\underline a \underline b, \, p}$ is valid for all diagonal metrics $\mathfrak{g}_{pq}$ whose components depend only on one of the coordinates describing the background spacetime, in our case $u$. 

Let us start introducing the set of vielbein\footnote{Here there is no understood summation over repeated indices.}
\be \label{vielbein}
e^{\underline i}=\sqrt{|\mathfrak{g}_{ii}(u)|} \, d\boldsymbol\varrho^i \, , \quad e^{\underline i}_{. \, j} = \delta^{\underline i}_{. \, j} \sqrt{|\mathfrak{g}_{ii}(u)|} \, , \quad \mathfrak{g}^{ii}(u) = (\mathfrak{g}_{ii}(u))^{-1} \,,
\ee
such that $d{\mathfrak s}^2 = \delta_{\underline p \underline q} e^{\underline p} e^{\underline q}$. The raising/lowering rules for the indices are
\be
e^{\underline a}_{. \, p} = \eta^{\underline a \underline b} \, e_{\underline b}^{. \, q} \, \mathfrak{g}_{pq}(u) \, , \quad e^{\underline a} (e_{\underline b}) = e^{\underline a}_{. \, p} e_{\underline b}^{. \, p}  = \delta^{\underline a}_{. \, \underline b}\,.
\ee
For the sake of simplicity, let us omit the dependence on $u$. From the above properties, we have
\be
e^{\underline c}_{. \, q} \, \partial_p e_{\underline b}^{\, . \, q} = \partial_p (e^{\underline c}_{. \, q} \, e_{\underline b}^{\, . \, q}) - e_{\underline b}^{\, . \, q} \, \partial_p e^{\underline c}_{. \, q} = \cancel{\partial_p (\delta^{\underline c}_{. \, b})} - \eta_{\underline b \underline d} \, \mathfrak{g}^{qr} e^{\underline d}_{\, . \, r} \, \partial_p e^{\underline c}_{. \, q} = - \eta_{\underline b \underline d} \, \mathfrak{g}^{qr} e^{\underline d}_{\, . \, r} \, \partial_p e^{\underline c}_{. \, q} \, .
\ee
Therefore, we can use the definition \eqref{spinconn} to write 
\be
\omega_{\underline a \underline b, \,p} = \eta_{\underline a \underline c} \, \omega^{\underline c}_{\, . \, \underline b, \,p} = \eta_{\underline a \underline c} \, \eta_{\underline b \underline d} \(- \mathfrak{g}^{qr} e^{\underline d}_{\, . \, r} \, \partial_p e^{\underline c}_{. \, q} + \mathfrak{g}^{qr} e^{\underline d}_{\, . \, r} \, e^{\underline c}_{. \, s} \, \Gamma^{s}_{pq}\) \, ,
\ee
where, recalling that $\mathfrak{g}_{pq}=\mathfrak{g}_{pq}(u)$,\footnote{Here $u$ is a fixed index related to the $u$-direction of the background spacetime.}
\be
\Gamma^{s}_{pq} = \frac{1}{2} \mathfrak{g}^{st} \( \partial_p \mathfrak{g}_{tq} + \partial_q \mathfrak{g}_{tp} - \partial_t \mathfrak{g}_{pq} \) = \frac{1}{2} \mathfrak{g}^{st} \( \delta^{u}_{. \, p} \partial_u \mathfrak{g}_{tq} + \delta^{u}_{. \, q} \partial_u \mathfrak{g}_{tp} \) - \frac{1}{2} \mathfrak{g}^{su} \partial_u \mathfrak{g}_{pq} \, .
\ee
Exploiting the diagonality of the metric and properties \eqnref{vielbein}, we get
\bea \label{spinconndef}
\omega_{\underline a \underline b, \,p} 
&= &\sum_{i,k} \eta_{\underline a \underline k} \, \eta_{\underline b \underline i} \(- \delta^{k}_{. \, i} \, \mathfrak{g}^{ii} e^{\underline i}_{\, . \, i} \, \partial_p e^{\underline i}_{. \, i} + \mathfrak{g}^{ii} e^{\underline i}_{\, . \, i} \, e^{\underline k}_{. \, k} \, \Gamma^{k}_{pi}\) = \nonumber \\
&& \sum_{i,k} \frac{\eta_{\underline a \underline k} \, \eta_{\underline b \underline i}}{\text{sgn}({\mathfrak{g_{ii}}})} \(- \frac{1}{2} \delta^{k}_{. \, i} \partial_p \ln \mathfrak{g}_{ii} + \sqrt{\frac{|\mathfrak{g}_{kk}|}{|\mathfrak{g}_{ii}|}} \,  \Gamma^{k}_{pi}\) \, .
\eea
Let us stress that in the above equation and in the following there are no understood summations. Now, if $p=u$
\begin{subequations}
  \begin{empheq}[left=\empheqlbrace]{align}
    &\Gamma^{k}_{ui}=\frac{1}{2} \, \delta^{k}_{. \, i} \, \mathfrak{g}^{ii} \partial_u \mathfrak{g}_{ii} = \frac{1}{2} \, \delta^{k}_{. \, i} \, \partial_u \ln \mathfrak{g}_{ii} \, ,  &\hspace{-50pt}  \text{if } i \neq u \, ,
    \\
    &\Gamma^{k}_{uu}=\delta^{k}_{. \, u} \, \mathfrak{g}^{uu} \partial_u \mathfrak{g}_{uu} - \frac{1}{2} \, \delta^{k}_{. \, u} \, \mathfrak{g}^{uu} \partial_u \mathfrak{g}_{uu} = \frac{1}{2} \, \delta^{k}_{. \, u} \, \partial_u \ln \mathfrak{g}_{uu} \, , & \hspace{-50pt} \text{if } i = u\, .
  \end{empheq}
\end{subequations}
All in all,
\be
\Gamma^{k}_{ui} = \frac{1}{2} \, \delta^{k}_{. \, i} \, \partial_u \ln \mathfrak{g}_{ii} \, , \quad \forall i , k\,.
\ee
So, from \eqnref{spinconndef}, it follows that
\be \label{spinconnu}
\omega_{\underline a \underline b, \, u} = 0 \, .
\ee
On the other hand, if $p \neq u$, then
\be
\partial_p \ln \mathfrak{g}_{ii} = \partial_p \ln \mathfrak{g}_{ii}(u) = 0 \, , \quad \forall i \, ,
\ee
and
\begin{subequations}
  \begin{empheq}[left=\empheqlbrace]{align}
    &\Gamma^{k}_{pi}=-\frac{1}{2} \, \delta^{k}_{. \, u} \, \delta^{i}_{. \, p} \, \mathfrak{g}^{uu} \partial_u \mathfrak{g}_{ii} \, ,  &\hspace{-200pt}  \text{if } i \neq u ,\\
  \nonumber  \\
    &\Gamma^{k}_{pu} = \frac{1}{2} \, \delta^{k}_{. \, p} \, \partial_u \ln \mathfrak{g}_{pp} \, , &\hspace{-200pt} \text{if } i = u\, .
  \end{empheq}
\end{subequations}
Therefore, plugging the above results in \eqnref{spinconndef}, we get
\be
\omega_{\underline a \underline b, \,p} 
= \sum_{i \neq u,k} \frac{\eta_{\underline a \underline k} \, \eta_{\underline b \underline i}}{\text{sgn}({\mathfrak{g_{ii}}})} \sqrt{\frac{|\mathfrak{g}_{kk}|}{|\mathfrak{g}_{ii}|}} \(-\frac{1}{2} \, \delta^{k}_{. \, u} \, \delta^{i}_{. \, p} \, \mathfrak{g}^{uu} \partial_u \mathfrak{g}_{ii}\) + \sum_{k} \frac{\eta_{\underline a \underline k} \, \eta_{\underline b \underline u}}{\text{sgn}({\mathfrak{g_{uu}}})} \sqrt{\frac{|\mathfrak{g}_{kk}|}{|\mathfrak{g}_{uu}|}} \(\frac{1}{2} \, \delta^{k}_{. \, p} \, \partial_u \ln \mathfrak{g}_{pp}\) \, ,
\ee
that is
\be
\omega_{\underline a \underline b, \,p} 
=\frac{1}{2}\, \text{sgn}(\mathfrak{g}_{uu}) \, \text{sgn}(\mathfrak{g}_{pp}) \, \frac{\partial_u \mathfrak{g}_{pp}}{\sqrt{|\mathfrak{g}_{uu}||\mathfrak{g}_{pp}}|} \(\eta_{\underline a \underline p} \, \eta_{\underline b \underline u}-\eta_{\underline a \underline u} \, \eta_{\underline b \underline p}\)  \, , \quad \forall p \neq u \, .
\ee

To sum up, given the set of vielbein in \eqnref{vielbein} and for all diagonal metrics whose components $\mathfrak{g}_{pq}$ depend only on a coordinate $u$, the components of the related spin-connection are
\be \label{spinconngen}
\omega_{\underline a \underline b, \,p} 
=\frac{1}{2} \, \text{sgn}(\mathfrak{g}_{uu}) \, \text{sgn}(\mathfrak{g}_{pp}) \, \frac{\partial_u \mathfrak{g}_{pp}(u)}{\sqrt{|\mathfrak{g}_{uu}(u)||\mathfrak{g}_{pp}(u)|}} \(\eta_{\underline a \underline p} \, \eta_{\underline b \underline u}-\eta_{\underline a \underline u} \, \eta_{\underline b \underline p}\)  , \quad \forall p \, .
\ee
In particular, this formula is valid for the Lorentzian Witten background given the approximation \eqnref{Omegaapprox}. Notice that this result agrees with the antisymmetric nature of $\omega_{\underline a \underline b, \,p}$ in $a \leftrightarrow b$ and reproduces \eqnref{spinconnu} for $p=u$.

\subsection{The Ramond-Ramond field strength} \label{appF4}

Now, let us turn our attention on the Ramond-Ramond field strength $F_4$. Given the approximation \eqnref{Omegaapprox}, from the definition in \eqnref{Wback} we have
\be
F_4=3 R^3 \omega_a \approx 3 R^3 dz^6 \wedge dz^7 \wedge dz^8 \wedge dz^9 \, .
\ee
Making use of the vielbein in \eqnref{vielbein}, we get
\be
F_4 = \frac{3 R^3}{(g_{zz}(u))^2} e^{\underline 6} \wedge e^{\underline 7} \wedge e^{\underline 8} \wedge e^{\underline 9} = \frac{3}{u} \, \varepsilon_{\underline a \underline b \underline c \underline d} \, e^{\underline a} \wedge e^{\underline b} \wedge e^{\underline c} \wedge e^{\underline d} \, , \quad \varepsilon_{\underline 6 \underline 7 \underline 8 \underline 9} = +1 \, .
\ee
Therefore,
\be \label{F4gen}
F^{(4)}_{\underline a \underline b \underline c \underline d} = \frac{3}{u} \, \varepsilon_{\underline a \underline b \underline c \underline d}
\, .
\ee

\subsection{The pull-back on the world-sheet of $D_p$ at the tip of the cigar} \label{appD}

The results of sections \ref{appspinconn} and \ref{appF4} can be exploited in order to compute the pull-back of \eqnref{D} at the tip of the cigar of the Lorentzian Witten background. In particular, from \eqnref{spinconngen} and \eqnref{F4gen} it follows that
\be
\omega_{\underline a \underline b, \,p} \Gamma^{\underline a \underline b}
= \text{sgn}(g_{pp}) \, \frac{\partial_u g_{pp}(u)}{\sqrt{g_{uu}(u)|g_{pp}(u)|}} \Gamma_{\underline p \underline u} \, ,
\ee
and
\be
F^{(4)}_{\underline a \underline b \underline c \underline d}\Gamma^{\underline a \underline b \underline c \underline d} =  \frac{3}{u} \, 4! \, \widetilde\Gamma \, , \quad \widetilde\Gamma = {\Gamma}^{\underline 6 \underline 7 \underline 8 \underline 9} \, .
\ee
All in all, using also the expression for the dilaton in \eqnref{Wback}, we get
\be \label{pullbackD}
D_\alpha =\partial_{\alpha} \rho^p D_p = \partial_\alpha + \sum_p \[\frac{1}{4} \, \text{sgn}(g_{pp}) \frac{\partial_{\alpha} \rho^p\partial_u g_{pp}(u)}{\sqrt{g_{uu}(u)|g_{pp}(u)|}} \Gamma_{\underline p \underline u} - \frac{3}{8} \frac{\partial_{\alpha} \rho^p \sqrt{|g_{pp}(u)|}}{(R^3 u)^{\sfrac{1}{4}}} \widetilde\Gamma \Gamma_{\underline p}\]   \, .
\ee
So, expanding around \eqnref{ansatz}, the above equation reduces to
\be
D_\alpha^{\text{tip}} = \partial_\alpha + \frac{1}{4} \frac{\partial_{\alpha} \Theta \partial_u g_{\theta\theta}(u_0)}{\sqrt{g_{uu}(u_0)g_{\theta\theta}(u_0)}} \Gamma_{\underline \theta \underline u} - \frac{3}{8} \frac{\partial_{\alpha} X^\mu \sqrt{g_{00}(u_0)}}{(R^3 u_0)^{\sfrac{1}{4}}} \widetilde\Gamma \Gamma_{\underline \mu} \, , \quad \mu = 0,1,...,d \, ,
\ee
where $ X^\mu$ are the classical $\mu$-components of $\rho$, for $\mu=0,1,...,d$. It follows that
\be
D_\alpha^{\text{tip}} = \partial_\alpha + \frac{1}{2} \partial_{\alpha} \Theta \Gamma_{\underline \theta \underline u} - \frac{3}{8} \partial_{\alpha} X^\mu \widetilde\Gamma \Gamma_{\underline \mu} 
\, .
\ee

\section{The generalized level-matching condition} \label{appctau}

In section \ref{secHagedorn}, we extended the Virasoro constraints \eqnref{classVirasoro} to the quantum word as \eqnref{quantumVirasoro}. Basically, \eqnref{classVirasoro} is a combination of \eqnref{Tabeq01} and \eqnref{Tabeq03}. So, we have not reported the details about the discussion of \eqnref{Tabeq02}. In this appendix, we show how to extract information from the quantum version of the latter condition.

Let us start from the bosonic sector and consider
\be \label{bosoffdiagconstr}
\int  d\tau d\sigma \, T^B_{\tau\sigma} = \int  d\tau d\sigma \, \partial_\tau \rho^p \partial_\sigma \rho^q \, g_{pq}(u) = \int  d\tau d\sigma \, \frac{1}{2} \(\sigma_1\)^{\alpha\beta} \partial_\alpha \rho^p \partial_\beta \rho^q \, g_{pq}(u) \, .
\ee
As usual, the above equation can be expanded around \eqnref{ansatzdef} up to quadratic order in quantum fluctuations. The result can be deduced by analogy from \eqnref{quantumVirasoro} through $\delta^{\alpha\beta} \mapsto \frac{1}{2} \(\sigma_1\)^{\alpha\beta}$. Since $\(\sigma_1\)^{\alpha\beta} \partial_\alpha X^\mu \partial_\beta X^\nu g_{\mu\nu}(u_0)=0$, we get
\be \label{quantum00}
\int  d\tau d\sigma \, T^B_{\tau\sigma} = \int d\tau d\sigma \{ \partial_\tau \widetilde \xi^p \partial_\sigma \widetilde \xi^q \, \delta_{pq} + \frac{9}{4} \, m_0^2 \, c_\tau \, c_\sigma \[(\widetilde \xi^u)^2+(\widetilde \xi^\theta)^2\] \} \, .
\ee
Given \eqnref{solKleinGordon}, let us compute explicitly every term of this expression. Starting from the first one for $p = u,\theta$, we have
\be
\begin{split}
\int_0^{2\pi} \hspace{-12pt} d\sigma \, \partial_\tau \widetilde \xi^p \partial_\sigma \widetilde \xi^p
= \pi\alpha' \sum_{n \ne 0} \frac{n}{\omega_n} \( \alpha_n^p \, \alpha_{-n}^p - \widetilde \alpha_{-n}^p \, \widetilde \alpha_n^p - \alpha_n^p \, \widetilde \alpha_n^p e^{-2 \, i \, \omega_n \, \tau} + \widetilde \alpha_{-n}^p \, \alpha_{-n}^p e^{+2 \, i \, \omega_n \, \tau} \)
\end{split} \, .
\ee
Then, exploiting \eqnref{brackets}, we get
\be \label{NmenoNtildequantumUTheta}
\begin{split}
\int_0^{2\pi} \hspace{-12pt} d\sigma \, \partial_\tau \widetilde \xi^p \partial_\sigma \widetilde \xi^p
&= 2\pi\alpha' \sum_{n > 0} \frac{n}{\omega_n} \( \alpha_{-n}^p \, \alpha_n^p - \widetilde \alpha_{-n}^p \, \widetilde \alpha_n^p \)
\end{split} \, , \quad p= u \, , \theta \, ,
\ee
and analogously
\be \label{NmenoNtildequantumOthers}
\begin{split}
\int_0^{2\pi} \hspace{-12pt} d\sigma \, \partial_\tau \widetilde \xi^q \partial_\sigma \widetilde \xi^q
&= 2\pi\alpha' \sum_{n > 0} \( \alpha_{-n}^q \, \alpha_n^q - \widetilde \alpha_{-n}^q \, \widetilde \alpha_n^q \)
\end{split} \, , \quad q= 0, 1, 2 , 3 , 6, 7 \, .
\ee

Let us stress that, given \eqnref{cancommrel},
\be
P_B = \frac{1}{2\pi\alpha'}  \int_0^{2\pi} \hspace{-12pt} d\sigma \, T^B_{\tau\sigma}
\ee
generates translations along the $\sigma$-direction in the bosonic sector, that is
\be
\[P_B, \widetilde \xi^q \] = - i \, \partial_\sigma \widetilde \xi^q \, .
\ee

Going forward, one can check that translations along the $\sigma$-direction in the fermionic sector are generated by
\be
P_F = +\frac{i}{4\pi\alpha'} \int_0^{2\pi} \hspace{-12pt} d\sigma \, \zeta_I^\kappa \partial_\sigma \zeta_{I'}^{\kappa'} \delta_{\kappa\kappa'} \delta^{II'} \, ,
\ee
that is
\be
\[P_F, \zeta_J^\kappa \] = - i \, \partial_\sigma \zeta_J^\kappa \, .
\ee
So, by analogy, we can extend the constraint in \eqnref{Tabeq02} as
\be
\mathcal{P} = P_B + P_F = 0 \, .
\ee

A similar computation to \eqnref{zeta1Dtauzeta1}, \eqnref{zeta2Dtauzeta2} and \eqnref{HFprime} leads to
\be \label{finalTFtausigma}
P_F = \sum_{r=\frac{1}{2}}^{+\infty} r \(S_{-r}^\kappa S_r^\kappa - \widetilde S_{-r}^\kappa \widetilde S_r^\kappa\) \, .
\ee
Notice that in \eqnref{NmenoNtildequantumUTheta}, \eqnref{NmenoNtildequantumOthers} and \eqnref{finalTFtausigma} all the creation operators are on the left and the annihilation ones on the right. This means that
\be
\langle 0 | \int_0^{2\pi} \hspace{-8pt} d\sigma \, \partial_\tau \widetilde \xi^p \partial_\sigma \widetilde \xi^p | 0 \rangle = 0 \, , \quad p=0,1,...,9 \, , \quad \langle 0 | P_F | 0 \rangle = 0 \, .
\ee
On the other hand, from \eqnref{intdsigmaxi2}, it follows that
\be
\langle 0 | \int_0^{2\pi} \hspace{-8pt} d\sigma \(\widetilde \xi^p\)^2 | 0 \rangle \ne 0 \, .
\ee
All in all, we conclude that
\be
\mathcal{P} = 0 \quad \Rightarrow \quad \langle 0 | \mathcal{P} | 0 \rangle = 0 \quad \Rightarrow \quad c_\tau c_\sigma = 0 \, .
\ee
Notice that the ansatz in \eqnref{ansatzdef} is written for real $c_\tau$ and $c_\sigma$ which can not be simultaneously equal to zero. Indeed, if so, $M_B$ would be zero and the mass-matching condition \eqnref{massmatchingcondition} would not be satisfied for sure. Moreover, if $c_\tau \ne 0$, $c_\sigma = 0$, then $M_B^2 =- \frac{9}{4} m_0^2 c_\tau^2 < 0$ and the spectrum would contain tachyons. The only possibility is thus
\be
c_\tau = 0 \, , \quad c_\sigma = \pm \frac{2}{3} \frac{1}{m_0} M_B = \pm \frac{M_B}{M_{KK}} \, .
\ee
Notice that, exploiting the above ansatz, the quantum version of the off-diagonal Virasoro constraint $\mathcal{P}$ reads
\be \label{generalizedlevelmatchcond}
\sum_{n=1}^{+\infty} \[\sum_{p=u,\theta} \frac{n}{\omega_n} \( \alpha_{-n}^p \, \alpha_n^p - \widetilde \alpha_{-n}^p \, \widetilde \alpha_n^p \) + \hspace{-4pt} \sum_{q=0,1,2,3,6,7} \hspace{-10pt} \( \alpha_{-n}^q \, \alpha_n^q - \widetilde \alpha_{-n}^q \, \widetilde \alpha_n^q \) \] + \sum_{r=\frac{1}{2}}^{+\infty} r \(S_{-r}^\kappa S_r^\kappa - \widetilde S_{-r}^\kappa \widetilde S_r^\kappa\) = 0\, .
\ee

Let us make a couple of observations. First of all, notice that the flat space limit (i.e., $\omega_n \to n$) of the above relation reproduces the usual level-matching condition $N_B+N_F=\widetilde N_B + \widetilde N_F$ for a closed string in flat space with winding modes but no quantized momentum in the $X^0$ direction.\footnote{Notice that, classically speaking, $p^0 \ne 0$ is forbidden by \eqnref{Tabeq02} for $h=1$.} 
 In other words, \eqnref{generalizedlevelmatchcond} is the generalized level-matching condition for closed strings embedded in the Witten background \eqnref{Wback} and localized at the tip of the cigar. Moreover, the left hand side of \eqnref{generalizedlevelmatchcond} is \emph{exactly} the world-sheet momentum in (2.5) of \cite{Hyun:2003ks}.
 
This discussion leads to the final proposal for the classical configuration adopted in this work, that is
\begin{subequations} \label{ansatzfinaldef}
  \begin{empheq}[left=\empheqlbrace]{align}
    & X^0 = m_0 \frac{\beta}{2\pi} \sigma \, , \quad X^0 \simeq X^0 + m_0\beta \, ,\\
    & X^i=\alpha' p^i m_0 \tau \, , \quad p^i \in \mathbb{R} \quad : \quad \widetilde \eta_{ij} \, p^i \, p^j = - M^2 \, , \quad i,j=1, ... , d \, ,\\
    & U=u_0 \quad \hspace{-8pt} : \quad \hspace{-8pt} \left . g_{uu}(U) \, \partial_\alpha U \right |_{U=u_0} \hspace{-4pt} = c_\alpha \, , \quad c_\tau = 0 \, , \, \, c_\sigma = \pm \frac{2}{3} \frac{1}{m_0} M_B = \pm \frac{M_B}{M_{KK}} \, , \qquad \quad \, \,  \,\\
    & \Theta=constant \, .
  \end{empheq}
\end{subequations}
\\ \,
\noindent Every parameter which initially appears in \eqnref{ansatz} has been fixed through an interplay between equations of motion, periodic boundary conditions for closed strings and the quantum version of the Virasoro constraints. The latter correspond to the generalized mass-shell condition \eqnref{massshellcond} and the generalized level-matching condition discussed in this appendix, that is \eqnref{generalizedlevelmatchcond}.

\newpage

\end{document}